\documentclass[twocolumn,times,tighten]{aastex63}

\usepackage{amsmath,amstext}
\usepackage{xcolor}

\newcommand{\iterations}{10,000 }

\usepackage[shortlabels]{enumitem}

\usepackage{ragged2e}
\usepackage{graphicx}
\usepackage{textcomp}
\usepackage{booktabs}
\usepackage{afterpage}
\usepackage{placeins}
\usepackage{amsmath}
\usepackage{multirow}
\usepackage[caption=false]{subfig}
\usepackage{tabularx}
\usepackage{scrextend}
\usepackage{hyperref}
\usepackage{verbatim}
\usepackage{lineno}

\shorttitle{Multi-wavelength Variability of Sgr A*}
\shortauthors{Boyce et al.}

\begin{document}

\title{Multi-wavelength Variability of Sagittarius A* in July 2019}

\author[0000-0002-6530-5783]{H. Boyce}
\affiliation{Department of Physics, McGill University, 3600 University Street, Montr{\'e}al, QC H3A 2T8, Canada}
\affiliation{McGill Space Institute, McGill University, 3550 University Street, Montr{\'e}al, QC H3A 2A7, Canada}

\author[0000-0001-6803-2138]{D. Haggard}
\affiliation{Department of Physics, McGill University, 3600 University Street, Montr{\'e}al, QC H3A 2T8, Canada}
\affiliation{McGill Space Institute, McGill University, 3550 University Street, Montr{\'e}al, QC H3A 2A7, Canada}

\author[0000-0003-2618-797X]{G. Witzel}
\affiliation{Max-Planck-Institut f\"ur Radioastronomie, Auf dem H\"ugel 69, D-53121, Bonn, Germany}

\author[0000-0002-9156-2249]{S. von Fellenberg}
\affiliation{Max-Planck-Institut f\"ur Radioastronomie, Auf dem H\"ugel 69, D-53121, Bonn, Germany}

\author[0000-0002-9895-5758]{S. P. Willner}
\affiliation{Center for Astrophysics \textbar\ Harvard \& Smithsonian, 60 Garden St., Cambridge, MA 02138-1516, USA}

\author{E. E. Becklin}
\affiliation{Department of Physics and Astronomy, University of California, Los Angeles, CA 90095-1547, USA}

\author[0000-0001-9554-6062]{T. Do}
\affiliation{UCLA Galactic Center Group, Physics and Astronomy Department, University of California, Los Angeles, CA 90024, USA}

\author[0000-0001-6049-3132]{A. Eckart}
\affiliation{Max-Planck-Institut f\"ur Radioastronomie, Auf dem H\"ugel 69, D-53121, Bonn, Germany}
\affiliation{Institute of Physics, University of Cologne, Z\"ulpicher Stra$\ss$e 77, 50937 Cologne, Germany}

\author[0000-0002-0670-0708]{G. G. Fazio}
\affiliation{Center for Astrophysics \textbar\ Harvard \& Smithsonian, 60 Garden St., Cambridge, MA 02138-1516, USA}

\author[0000-0003-0685-3621]{M. A. Gurwell}
\affiliation{Center for Astrophysics \textbar\ Harvard \& Smithsonian, 60 Garden St., Cambridge, MA 02138-1516, USA}

\author[0000-0002-5599-4650]{J. L. Hora}
\affiliation{Center for Astrophysics \textbar\ Harvard \& Smithsonian, 60 Garden St., Cambridge, MA 02138-1516, USA}

\author[0000-0001-9564-0876]{S. Markoff}
\affiliation{Anton Pannekoek Institute for Astronomy, University of Amsterdam, Science Park 904, 1098 XH Amsterdam, The Netherlands}
\affiliation{Gravitation Astroparticle Physics Amsterdam (GRAPPA) Institute, University of Amsterdam, Science Park 904, 1098 XH Amsterdam, The Netherlands}

\author[0000-0002-6753-2066]{M. R. Morris}
\affiliation{Department of Physics and Astronomy, University of California, Los Angeles, CA 90095-1547, USA}

\author[0000-0002-8247-786X]{J. Neilsen}
\affiliation{Villanova University, Mendel Science Center Rm. 263B, 800 E Lancaster Ave, Villanova PA 19085, USA}

\author[0000-0001-6923-1315]{M. Nowak}
\affiliation{Physics Department, Washington University CB 1105, St Louis, MO 63130, USA}

\author{H. A. Smith}
\affiliation{Center for Astrophysics \textbar\ Harvard \& Smithsonian, 60 Garden St., Cambridge, MA 02138-1516, USA}

\author[0000-0002-2967-790X]{S. Zhang}
\affiliation{Bard College, 30 Campus Road, Annandale-on-Hudson, New York, 12504, USA}

\correspondingauthor{H. Boyce}
\email{hope.boyce@mail.mcgill.ca}

\begin{abstract}
We report timing analysis of near-infrared (NIR), X-ray, and sub-millimeter (submm) data during a three-day coordinated campaign observing Sagittarius A*. Data were collected at 4.5$\mu$m with the Spitzer Space Telescope, $2-8$ keV with the Chandra X-ray Observatory, $3-70$ keV with NuSTAR, 340 GHz with ALMA, and at 2.2$\mu$m with the GRAVITY instrument on the Very Large Telescope Interferometer. Two dates show moderate variability with no significant lags between the submm and the infrared at 99\% confidence. July 18 captured a moderately bright NIR flare (F$_{\mathrm{K}}\sim$ 15 mJy) simultaneous with an X-ray flare (F$_{2-10 \mathrm{keV}}\sim$ 0.1 cts/s) that most likely preceded bright submm flux (F$_{340 \mathrm{GHz}}\sim$ 5.5 Jy) by about $+34\substack{+14\\ -33}$ minutes at 99\% confidence. The uncertainty in this lag is dominated by the fact that we did not observe the peak of the submm emission. A synchrotron source cooled through adiabatic expansion can describe a rise in the submm once the synchrotron-self-Compton NIR and X-ray peaks have faded. This model predicts high GHz and THz fluxes at the time of the NIR/X-ray peak and electron densities well above those implied from average accretion rates for Sgr A*. However, the higher electron density postulated in this scenario would be in agreement with the idea that 2019 was an extraordinary epoch with a heightened accretion rate. Since the NIR and X-ray peaks can also be fit by a non-thermal synchrotron source with lower electron densities, we cannot rule out an unrelated chance coincidence of this bright submm flare with the NIR/X-ray emission. 
\end{abstract}


\section{Introduction}

Sagittarius A* (Sgr A*), the supermassive black hole (SMBH) sitting at the bottom of the central gravitational potential of the Milky Way, co-exists in a dynamic environment with a dense stellar cluster, active star formation, and hot, inefficiently accreting gas. Over the past $\sim$20 years, the mass ($\sim4\times10^{6}$M$_{\sun}$) and accretion rate ($\lesssim10^{-7}$M$_{\sun}$ yr$^{-1}$) of Sgr A* have been pinned down through careful analysis of stellar orbits \cite[e.g.,][]{2016ApJ...830...17B,2017ApJ...837...30G} and multi-wavelength flux measurements \cite[e.g.,][]{2003ApJ...591..891B,2006ApJ...640..308M,2007ApJ...654L..57M,2012ApJ...755..133S, 2015ApJ...809...10Y,Bower+2018}. These properties, along with its low bolometric-to-Eddington luminosity ratio \citep[L/L$_{\mathrm{Edd}}\sim$10$^{-9}$;][]{2010RvMP...82.3121G} and characterizations of the quiescent spectral energy distribution (SED) have motivated models of advective and inefficient accretion flows \citep[e.g.,][]{2002ApJ...575..855Q, 2003ApJ...591..891B, 2003ApJ...598..301Y, 2004ApJ...611L.101L, 2006ApJ...640..319X, 2013Sci...341..981W}.

Though Sgr A* seems to be variable at every wavelength it has been observed, the physical mechanisms behind the changes in Sgr A*'s flux density remain uncertain. Physical models often invoke populations of accelerated electrons caused by magnetic re-connection events, jets, sudden disk instabilities, or other stochastic processes in the accretion flow \citep[e.g.,][]{2001A&A...379L..13M, 2002ApJ...566L..77L, 2003ApJ...598..301Y, 2004ApJ...611L.101L,2009ApJ...703L.142D,2009A&A...508L..13M,2010ApJ...725..450D,2016ApJ...826...77B,2017MNRAS.468.2552L}.  Additional models attempt to explain the variability in the context of tidal disruption of asteroids \citep{2008A&A...487..527C, 2009A&A...496..307K, 2012MNRAS.421.1315Z} or expanding plasma blobs \citep[e.g.,][]{1966Natur.211.1131V, 2006ApJ...650..189Y,2008ApJ...682..373M, 2015MNRAS.454.3283Y, 2017MNRAS.468.2552L}.  Finally, emission may be amplified through strong gravitational lensing near the event horizon \citep[e.g.,][]{2015ApJ...812..103C}.

Variability in the NIR is expected to arise from a fluctuating non-thermal population of electrons. \cite{2019ApJ...882L..28C} showed that Sgr A*'s IR variability was statistically consistent over two decades, never deviating from log-normal distribution of flux densities. This consistency was immediately challenged by the unprecedented IR brightness detected by \cite{2019ApJ...882L..27D} in March 2019. Such a deviation from the usual statistical behaviour \citep{2009ApJ...691.1021D,2009ApJ...694L..87M,2011ApJ...728...37D,2012ApJS..203...18W,2014ApJ...793..120H,2018ApJ...863...15W} challenges the current model and could point to a dynamical interaction or a temporary change in Sgr A*'s accretion state.

The submm-THz bump in Sgr A*'s SED is often attributed to a steady synchrotron source originating from a thermal electron distribution \citep[e.g.,][]{Bower+2018,2018ApJ...862..129V}. Submm flux from Sgr A* is variable down to timescales of seconds to hours \citep{2021ApJ...920L...7M,2020ApJ...892L..30I}, suggesting source sizes on the same order as the BH's innermost stable orbit. \cite{2014MNRAS.442.2797D} found an 8-hour characteristic timescale for the variability by analyzing light curves over a period of 10 years. \cite{2017A&A...601A..80S} presented a statistical analysis of submm variability at 345 GHz from 2008 to 2014, reporting a mean flux density measurement of $\sim$3 Jy. In 190 hours of observations, the 345 GHz flux rose above 4.5 Jy only four times. \cite{2021ApJ...920L...7M} reported observations of Sgr A* at 230 GHz in June 2019, finding that the mean flux level was 3.74 Jy: 20\% higher than in 2015-2017 and 3\% higher than levels in 2009-2012 and 2013-2014. Such variability (on the scale of $\sim$10 years) is similar to the expected global mass accretion variability \citep{2020MNRAS.492.3272R}. 

Sgr A*'s faint, steady X-ray emission \citep{2001Natur.413...45B,2003ApJ...591..891B} is most likely thermal bremsstrahlung emission originating in the accretion flow near the Bondi radius \citep{2002ApJ...575..855Q, 2003ApJ...591..891B, 2003ApJ...598..301Y, 2004ApJ...611L.101L, 2006ApJ...640..319X, 2013Sci...341..981W}. This quiescent state is interrupted about once per day by distinct X-ray flares of non-thermal emission presumed to be coming from very close to the black hole \citep{2013ApJ...774...42N, 2015ApJ...799..199N,2019ApJ...886...96H,2017ApJ...843...96Z}. The flux density distribution of the X-ray variability can be described by a power law \citep[e.g.,][]{2015ApJ...799..199N} or log-normal \citep{witzel+2021}. Recent examination of long term X-ray variability suggests that Sgr A*'s flaring rate can change over the span of several years \citep{2022MNRAS.510.2851A}.

There have been several studies reporting a correlation between submm and NIR/X-ray variability \citep[e.g.,][]{2006A&A...450..535E,2006ApJ...644..198Y,2008A&A...492..337E,2009ApJ...706..348Y,2011A&A...528A.140T,2012A&A...537A..52E,2016A&A...589A.116M,2018ApJ...864...58F}. These provide increasing evidence that the submm and NIR/X-ray sources are physically or radiatively connected. Correlations between the radio and NIR remain less clear \citep{2017ApJ...845...35C}.

To connect physical models with observables, studies have analyzed both the timing properties between wavelengths and SED characteristics of Sgr A* during quiescence and flares. They aim to put constraints on what radiative mechanisms must be at play. For example, there are models that predict simultaneity of NIR/X-ray flares through synchrotron self-Compton (SSC) processes \citep{2001A&A...379L..13M,2008A&A...479..625E}, those that cool the electrons of the synchrotron source to predict delayed low-frequency emission relative to the NIR/X-ray \citep[e.g.,][]{2006ApJ...644..198Y,witzel+2021}, and those that connect time lags to relativistic outflows \citep[e.g.,][]{2021arXiv210713402B}. General-relativistic magneto-hydrodynamic (GRMHD) simulations also predict radiative models and observable SED characteristics scaled to Sgr A* \citep[e.g.,][]{2009ApJ...706..497M,2014A&A...570A...7M}, and even simulate light curves comparable to observations \citep{2021MNRAS.507.5281C}.

There are several observational avenues that can be used to constrain properties of the plasma in the galactic centre. Observations of a magnetar at an angular distance of $\sim2.5$ arcsec from Sgr A* \citep{2013ApJ...770L..23M,2013ApJ...775L..34R} have been useful in constraining the interstellar scattering that affects observations in the vicinity of the SMBH \citep[e.g.][]{2015ApJ...798..120B,2017MNRAS.471.3563D}. Such observations can constrain properties of the plasma and magnetic field \citep[e.g.][]{2013Natur.501..391E}. Even closer to the black hole, new observations by the Event Horizon Telescope (EHT) \citep[e.g.,]{2021ApJ...915...99I} and GRAVITY are beginning to probe the plasma and general relativistic effects near the event horizon. EHT observations of Sgr A* were collected in 2017, 2018, and are scheduled for 2022. Such high-resolution imaging will help untangle the dynamics of the plasma immediately around Sgr A* from the significant interstellar scattering between earth and the Galactic Center \citep[e.g.,][]{2018ApJ...865..104J,2019A&A...629A..32I}.  Also probing near event-horizon scales, the GRAVITY Collaboration has demonstrated that exceptionally precise near-infrared interferometry of Sgr A*'s position can probe the apparent motion of its centroid. This in turn can be successfully modelled as a hot-spot orbiting less than 10 gravitational radii away from the SMBH \citep{2018A&A...618L..10G}.

\begin{table*}[t]
    \caption{Data sets analysed in this work.}
    \label{tab:observations}
    \tabletypesize{\small}
\begin{tabularx}{0.98\textwidth}{lccccccccc}
	
	\toprule
	Observatory & Date & OBSID & Start & End & Energy & Wavelength & $\#$Ant & Baselines & calibrators \\
	 & (UT) &  & (UT) & (UT) &  & [frequency] & & (k$\lambda$)& \\
	\toprule
	NuSTAR & 2019-07-17 & 30502006002 & 21:51:09 & 08:34:21 & 3$-$70 keV & 6.2$-$0.2 \AA& - & - & - \\
	 & 2019-07-26 & 30502006004 & 00:41:09 & 10:21:06 & 3$-$70 keV & 6.2$-$0.2 \AA& - & - & - \\
	\midrule
	Chandra & 2019-07-17 & 22230 & 22:51:26 & 14:51:26 & 2$-$8 keV & 6.2$-$1.6 \AA & - & - & - \\
	 & 2019-07-21 & 20446 & 00:00:14 & 16:00:14 & 2$-$8 keV & 6.2$-$1.6 \AA & - & - & - \\
	 & 2019-07-26 & 20447 & 01:32:40 & 17:32:40 & 2$-$8 keV & 6.2$-$1.6 \AA & - & - & - \\
	\midrule
	GRAVITY & 2019-07-17 & 0103.B-0032(D) & 23:32:55 & 05:32:55 & 0.7$-$0.8 eV & 2.2$-$1.65 $\mu$m & - & - & - \\
	\midrule
	Spitzer & 2019-07-17 & 69965312 & 23:21:33 & 07:21:20 & 0.3 eV & 4.5 $\mu$m & - & - & - \\
	 & 2019-07-18 & 69965568 & 07:25:02 & 15:24:49 & 0.3 eV & 4.5 $\mu$m & - & - & - \\
	 & 2019-07-21 & 69965824 & 00:21:47 & 08:21:37 & 0.3 eV & 4.5 $\mu$m & - & - & - \\
	 & 2019-07-21 & 69966080 & 08:24:49 & 16:25:05 & 0.3 eV & 4.5 $\mu$m & - & - & - \\
	 & 2019-07-26 & 69966336 & 02:02:35 & 10:02:22 & 0.3 eV & 4.5 $\mu$m & - & - & - \\
	 & 2019-07-26 & 69966592 & 10:06:02 & 18:05:53 & 0.3 eV & 4.5 $\mu$m & - & - & - \\
	\midrule
	ALMA & 2019-07-17 & 2018.A.00050.T & 23:49:02 & 06:49:56 & 0.0014 eV & [340 GHz] & 11 & 10.1$-$54.4 & J1700-2610 \\
	 &  &  &  & & & & &  & J1733-3722 \\
	 & 2019-07-20 & 2018.A.00050.T & 03:55:59 & 06:47:57 & 0.0014 eV & [340 GHz] & 11 & 10.1$-$54.4 & J1700-2610 \\
	 & 2019-07-25 & 2018.A.00050.T & 23:51:49 & 06:45:15 & 0.0014 eV & [340 GHz] & 10 & 10.1$-$54.4 & J1717-3342 \\
	\bottomrule
	
\end{tabularx}

\end{table*}

Numerous joint X-ray and IR campaigns have observed Sgr A* over the last 16 years \citep{2004A&A...427....1E,2006A&A...450..535E,2006ApJ...644..198Y,2008A&A...479..625E,2009ApJ...698..676D,2009ApJ...706..348Y,2012AJ....144....1Y,2016A&A...589A.116M,2018ApJ...864...58F,2017MNRAS.468.2447P}. Our joint Spitzer and Chandra study reported $\sim$144 hours of coordinated observations collected between 2014 and 2017 \citep{2019ApJ...871..161B}. These observations captured four modestly bright multi-wavelength flares from Sgr A*. Comparing the X-ray observations to simulations of the infrared statistical behaviour \citep{2018ApJ...863...15W}, the consistent observation of X-ray and IR events within 20 minutes of each-other point to a physical connection between the emission at these wavelengths, rather than chance association. In \cite{2019ApJ...871..161B} we found the time lag between the peaks in the X-rays and the peaks in the IR was consistent with simultaneity and at most on order of 10$\sim$20 minutes.

Here we extend our original study by investigating the physical and temporal correlations between X-ray and IR variability with Spitzer and Chandra observations of Sgr A* in the summer of 2019, alongside simultaneous NuSTAR, GRAVITY, and ALMA monitoring. To constrain the particle acceration responsible for flaring, \cite{seb_flare_paper} analyze the Spitzer, GRAVITY, NuSTAR, and Chandra data of July 17$-$18 in the context of time-resolved SED modelling and found that the NIR and X-ray flare can be best modelled with a non-thermal synchrotron source. \cite{Michail+2021} combine the Spitzer NIR measurements with the 340 GHz ALMA measurements on July 18 to explore models that describe the NIR as SSC of a synchrotron source responsible for delayed submm emission adiabatically expanding. Bringing all available data together, this paper reports timing analysis between the five observatories on July 17$-$18, July 21, and July 26, and explores how SED models (see Section \ref{Discussion}) can be constrained by the submm, NIR, and X-ray timing data.

\vspace{2cm}
\section{Observations and Data Reduction} \label{Observations}
\begin{figure*}
\plotone{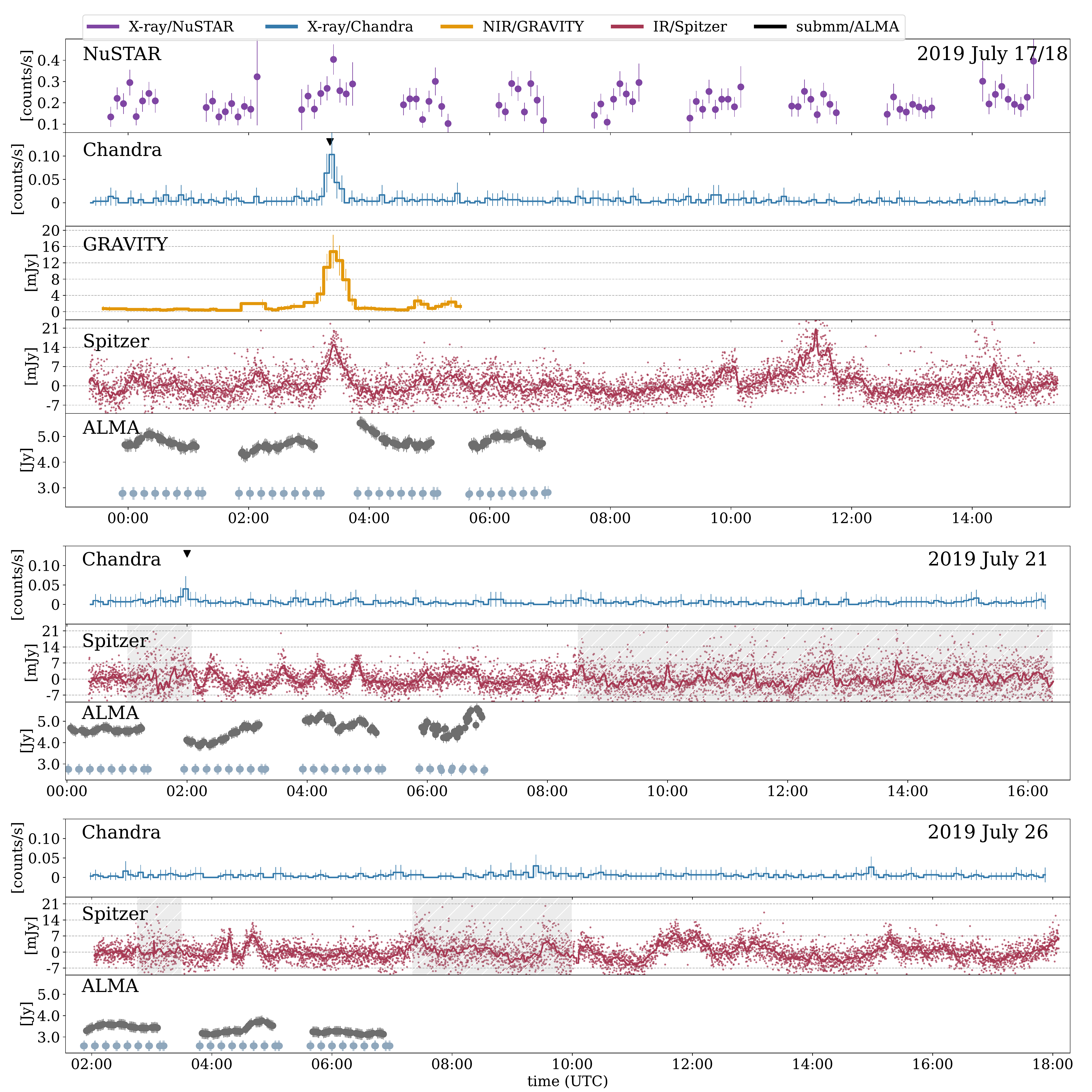}
\caption{Simultaneous submm, IR, and X-ray light curves of Sgr A* from July 2019. The purple, blue, orange, red, and grey data show the NuSTAR 3--70 keV, Chandra 2--8 keV, GRAVITY 2.2 $\mu$m, Spitzer 4.5 $\mu$m, and ALMA 340 GHz data, respectively. The x-axis displays the barycentre-corrected UT on each date. Spitzer data at 4.5$\mu$m is the excess flux density (mJy) of the pixel containing Sgr A*  \citep[see \S 2.1 of][]{2018ApJ...863...15W} offset with a value of 1.9 mJy and de-reddened with the extinction A$_{\mathrm{Ks}}=2.42\pm0.002$ from \cite{2011ApJ...737...73F}. The red line shows the data binned over 3.5 minutes. Grey regions on July 21 and 26 indicate where the light curve is unreliable due to decreased signal-to-noise. The GRAVITY light curve was de-reddened assuming a K-band extinction of $2.42\pm0.01$ magnitudes. Significant X-ray flares in Chandra data are identified by the Bayesian Blocks routine (p$_{0}=0.05$) and indicated here with black arrows. ALMA calibrators are plotted in light grey for comparison.}
\label{fig:lightcurves}
\end{figure*}

The IRAC instrument \citep{2004ApJS..154...10F} on the Spitzer Space Telescope \citep{2004ApJS..154....1W} observed Sgr A* at 4.5 $\mu$m for eight $\sim$24-hour-long stretches between 2013 and 2017. Six of these observations had simultaneous monitoring from the Chandra X-ray Observatory \citep{2000SPIE.4012....2W} and are reported by \cite{2019ApJ...871..161B}. Since then, three additional epochs of simultaneous monitoring totalling $\sim$48-hours were observed. These additional epochs expand the total dataset to $\sim$155 hours of simultaneous X-ray and IR data.  Figure \ref{fig:lightcurves} displays these new 2019 epochs along with additional coordinated coverage from NuSTAR, GRAVITY, and ALMA. For an assumed distance of 8.2 kpc, $1^{\prime\prime}=0.040$ pc \citep{GRAVITY_distance} \footnote{7.9 kpc would give $1^{\prime\prime}=0.038$ pc \citep{UCLA_distance}}.

\subsection{Spitzer}

All Spitzer observations were collected in a similar manner as the previous epochs in the campaign. \cite{2014ApJ...793..120H} give a complete description of the Sgr A* Spitzer monitoring campaign. We offer a brief summary here. Three observing blocks were collected at 4.5 $\mu$m (filter width of 1$\mu$m) in each of the three 16 hour epochs: an initial mapping operation performed after the slew to the Sgr A* field followed by two successive 8 hour staring operations. Each staring operation began by using the ``PCRS Peakup" mode to position Sgr A* on the center of pixel (16,16) in the IRAC subarray. The subarray mode for Spitzer/IRAC reads out 64 consecutive images (a ``frame set") of a 32x32 pixel region on the IRAC detector. This frame set is known as one Basic Calibrated Data product (BCD), which is the data format downloaded from the Spitzer Heritage Archive\footnote{The Spitzer Heritage Archive (\url{http://irsa.ipac.caltech.edu}) is part of the NASA/ IPAC Infrared Science Archive, which is operated by the Jet Propulsion Laboratory, California Institute of Technology, under contract with the National Aeronautics and Space Administration.}. Each component in the frame set is a 0.1 s 32$\times$32 image, so one frame set takes 6.4 s to complete. After converting the pixel intensity into mJy, each frame set was combined into a single 32$\times$32 image referred to as a ``6.4 s BCD coadd". Consecutive frame sets were typically separated by 2 s of telescope overheads, and this resulted in an observation cadence of approximately 8.4 s per frame. 

To extract light curves of Sgr A* from the Spitzer/IRAC data, we used the same methodology as \cite{2019ApJ...871..161B} and \cite{2018ApJ...863...15W}, including an updated procedure based on the steps described in Appendix A1 of \cite{2014ApJ...793..120H}. This procedure corrects for the varying intra-pixel sensitivity of the Spitzer/IRAC detector and the effect of nearby sources on the measured flux of Sgr A* as the telescope pointing jitters during the observations. The resulting light curves are the excess variable flux density in pixel (16,16) measured relative to the non-variable stellar background ($\sim$ 250 mJy). The baseline flux density of these IR light curves is unknown, though the value has been inferred to be 1.9 mJy from the cumulative distributions of flux densities of Sgr A* \citep{2018ApJ...863...15W}. As in \cite{seb_flare_paper}, we added an offset of 1.9 mJy and de-reddened the resulting values with the extinction A$_{\mathrm{Ks}}=2.42\pm0.002$ from \cite{2011ApJ...737...73F} to produce the light curves plotted in Figure \ref{fig:lightcurves}.

\subsection{Chandra}

The simultaneous Chandra observations were acquired using the ACIS-S3 chip in the FAINT mode with a 1/8 subarray. The small subarray was chosen to avoid photon pileup during bright flares from Sgr A* and the nearby magnetar, SGR J1745$-$2900 \citep{2013ApJ...770L..23M,2013ApJ...775L..34R,2015MNRAS.449.2685C, 2017MNRAS.471.1819C}. 

\par We performed Chandra data reduction and analysis with CIAO v4.9 tools\footnote{Chandra Interactive Analysis of Observations (CIAO) software is available at http://cxc.harvard.edu/ciao/} \citep{2006SPIE.6270E..1VF} and calibration database 4.7.3. The \texttt{chandra\_repro} script was used to reprocess level 2 events files before the WCS coordinate system was updated (\texttt{wcs\_update}). Barycentric corrections to the event times were performed with the CIAO tool \texttt{axbary}. We extracted a 2--8 keV light curve from a circular region of radius 1.25$^{\prime\prime}$ centered on Sgr A*. The small extraction region and energy range isolate Sgr A*'s emission from the nearby magnetar \citep[e.g.,][]{2013ApJ...770L..23M,2013ApJ...775L..34R,2017MNRAS.471.1819C} and from the diffuse X-ray background \citep[e.g.,][]{2003ApJ...591..891B, 2012ApJ...759...95N,2013Sci...341..981W}. X-ray light curves are plotted in purple in Figure \ref{fig:lightcurves}. Flux densities for SED modelling (Section \ref{Discussion}) were corrected for dust scattering and absorption as described in \cite{seb_flare_paper}.

\subsection{NuSTAR}

The NuSTAR \citep{2013ApJ...770..103H} data have been processed using the NuSTAR Data Analysis Software NUSTARDAS, HEASOFT v. 6.28, and CALDB v20200912. Data were filtered for periods of high instrumental background due to South Atlantic Anomaly passages and known bad detector pixels.
We analysed the observations starting on July 17, 2019 21:51:09 and on July 26, 2019 00:41:09 (ObsID: 30502006002 and 30502006004, respectively). We applied the barycenter corrections. 
Light curves and spectra were extracted via the \texttt{nuproducts} tool from a region of radius 20$^{\prime\prime}$ centered on the position of Sgr A*. Because the focal plane modual B (FPMB) is contaminated by stray light from faraway bright X-ray sources outside of the field of view, we only present the analysis of the FPMA data (the results obtained with FPMB are consistent with the results). The light curves were accumulated in the 3$-$10 keV band and with 380 s time bins for comparison with the GRAVITY data. Bins with small fractional exposures were removed. Flux densities for SED modelling (Section \ref{Discussion}) were corrected for dust scattering and absorption as described in \cite{seb_flare_paper}.

\subsection{GRAVITY}

The K-band (2.1$-$2.4 $\mu$m) GRAVITY light curve was derived from the coherent flux measurement of Sgr A* as described by \cite{flux_distribution_paper} and \cite{seb_flare_paper}. We derived the flux ratios relative to S2 using separate observations. We de-reddened the flux assuming a K-band extinction of $2.42\pm0.01$ magnitudes. The light curve has been corrected for the contamination of S2 at the edge of the field of view and errors were scaled in the same way as described in \cite{flux_distribution_paper}. We ignored the contribution of the faint star S62 \citep{faint_star_paper}, which should amount to less than $0.1~\mathrm{mJy}$. The H-band light curve was also reduced but not used here as the lower signal-to-noise provided negligible improvement over the K-band data in constraining the timing. See \cite{seb_flare_paper} for details.

\subsection{ALMA}

All three epochs of Spitzer data presented here were partly covered by ALMA observations\footnote{project 2018.A.00050.T, PI: J.Carpenter}.
Sgr~A* was observed using the 7m ALMA compact array July 17/18 \citep[see also][]{Michail+2021}, 21, and 26 in 2019. With eleven and ten (epoch of July 25/26) antennas, this compact configuration has fifty-five and fourty-five unique projected baselines, respectively, from 8.904 m to 47.987 m (10.1 to 54.4 k$\lambda$). The corresponding maximal resolution is 4.6$^{\prime\prime}$. The total continuum bandwidth was 2 GHz.

The quality assessment of the epochs by the ALMA pipeline was ``semi-pass’’ for the first two epochs and ``pass’’ for the last epoch\footnote{Criteria described in the ALMA technical handbook \url{https://almascience.nrao.edu/documents-and-tools/cycle7/alma-technical-handbook/view}}. Each epoch consisted of four observation blocks on Sgr A*, each $\sim$76min, with  seven scans of $\sim$7min duration and an eighth scan that is shorter than 1min. Between each scan there is a gap of $\sim$4min, and between the observation blocks there are gaps of $\sim$40min. The data quality particularly suffered from the atmospheric conditions in the last observing block of each of the first two epochs, while all other blocks are of comparable quality.

Bandpass and gain were calibrated using calibrators J1337$-$1257 (block 1 of each epoch) and J1924$-$2914 (blocks 2$-$4 of each epoch). Gain and phase calibration were executed using the calibrators J1700$-$2610 (epoch 1, blocks 1$-$3; epoch 2), J1733$-$3722 (epoch 1, block 4), and J1717$-$3342 (epoch 3) in alternation with measurements of Sgr A*.

To derive light curves we first restored the gain-calibrated visibilities with the scripts \texttt{scriptForPi.py} which are part of the data archive. The resulting visibilities were then separated by source and spectral range. For each spectral window with science data (16, 18, 20, and 22), we chose the frequency range dominated by continuum emission as identified by the routine \texttt{hif\_findcont} of the ALMA pipeline. We then applied three iterations of  fitting a point source model to the visibilities (with the CASA routine \texttt{uvmodelfit}) and interleaved phase self-calibration (with the CASA routines \texttt{gaincal} and \texttt{applycal}). After a fourth fit with a point source model, we used the resulting flux density as our measurement. This algorithm was applied to visibilities of Sgr A* and the particular phase calibrator in time windows of 1 minute. The last 1-minute bin of each scan with just a few datapoints, as well as the last scans of observations blocks that are shorter than 1 minute, were discarded.

The resulting light curves have a regular cadence of 1 min and a total duration of ~7 hours with 5 hours of data each. Heliocentric corrections of +7.366 min, +7.158 min, and +6.772 min were applied for the comparison with the Spitzer light curves. We estimate the absolute flux density calibration to be accurate within 10$\%$ uncertainty and the relative photometric precision to be $<3\%$.

\section{Analysis} \label{Analysis}

\subsection{Flare Characterization}\label{FlareAnalysis}
To identify significant X-ray flares, we used the Bayesian Blocks algorithm as described by \cite{1998ApJ...504..405S} and \cite{2013ApJ...764..167S} and provided as a python routine by Peter K. G. Williams \citep[\texttt{bblocks};][]{2017ascl.soft04001W}. We ran the algorithm using a 95\% confidence interval (a false positive rate of $p_{0}=0.05$). This choice for $p_{0}$ implies that the probability that a change point is real is $1-0.05 = 95$\%, and the probability that a flare (at least two change points) is real is $1-(p_{0})^{2}=99.8\%$. Detected flares are indicated by triangles in Figure \ref{fig:lightcurves}.
\par We detected two Chandra X-ray flares during the total overlap-period of X-ray and IR, one on 2019 July 18 and one on 2019 July 21. The detection rate is consistent with past measurements of the average number of X-ray flares from Sgr A* \citep[$\sim$1.1/day;][]{2015ApJ...799..199N,2015MNRAS.454.1525P}.  The mean quiescent flux measured with Chandra during these epochs was 0.005 counts-per-second (cts/s), and while the flare detected on 2019 July 21 was similar to those reported by \cite{2019ApJ...871..161B}; (20 counts), the flare detected on July 18, 2019 had a total of 74 counts and was not bright enough for pile-up to significantly affect the measurement.

\par In contrast to the distinct peaks in the X-rays, the emission from Sgr A* at IR wavelengths is constantly varying. An apparent quasi-periodic feature appears in the Spitzer light curve on July 21. Such apparent periodicities can appear in processes described by correlated red-noise and the statistics of Sgr A*'s NIR variability is well described by a red-noise process \citep[e.g.][]{2009ApJ...691.1021D,2012ApJS..203...18W}. There are also multiple IR peaks where we see no significant X-ray emission, even in cases when the IR emission is most elevated (e.g., $\sim$6 mJy around 11:30 July 18), whereas the  X-ray flare on July 18 was accompanied by a significant rise in the NIR flux density levels. This behaviour (NIR peaks accompanying X-ray flares but not the reverse) is consistent with all previous reported X-ray/IR observations of Sgr A* as well as recent simulations \citep[e.g.,][]{witzel+2021}. We do not consider the X-ray flare with a lack of NIR rise around 02:00 July 21 as contradictory because the IRAC data exhibited higher-than-normal noise levels at this time due to poor stability in the telescope pointing. A rise in the submm flux at 06:30 on July 21 was not accompanied by corresponding variability X-ray, and has marginally significant higher-than average variability in the NIR. Additionally, on July 26 IR variability was observed along with a rise in the submm but with no corresponding flare in the X-ray.

\vspace{5mm}

\par With a K-band peak flux density of $\sim$16 mJy, the NIR flare on July 18 can be classified as moderately bright in the context of previously observed variability \citep{seb_flare_paper} while the X-ray flare was fairly modest with a peak of 0.1 cts/s. This is a factor of $\sim$2 brighter than the four faint flares with simultaneous Spitzer data reported by \cite{2019ApJ...871..161B}, but a factor of $\sim$14 lower than the brightest X-ray flare observed \citep{2019ApJ...886...96H}. The brightest flare observed simultaneously in NIR and X-rays was reported by \cite{2009ApJ...698..676D}, and had an L-band flux density of $\sim$25 mJy and the 2-10 keV X-ray flare reaching $\sim$1 cts/s. While the X-ray and NIR variability was moderate on July 18, this does not hold for 340 GHz, which at the highest point was 5.5 Jy, well above the typically measured quiescent levels of $\sim$3 Jy \citep{2017A&A...601A..80S}. In fact, the mean flux density ($\sim$4.5 Jy) measured on July 18 and July 21, was also elevated with respect to historic levels.

\begin{figure*}[ht!]
\centering
\includegraphics[width=0.9\textwidth]{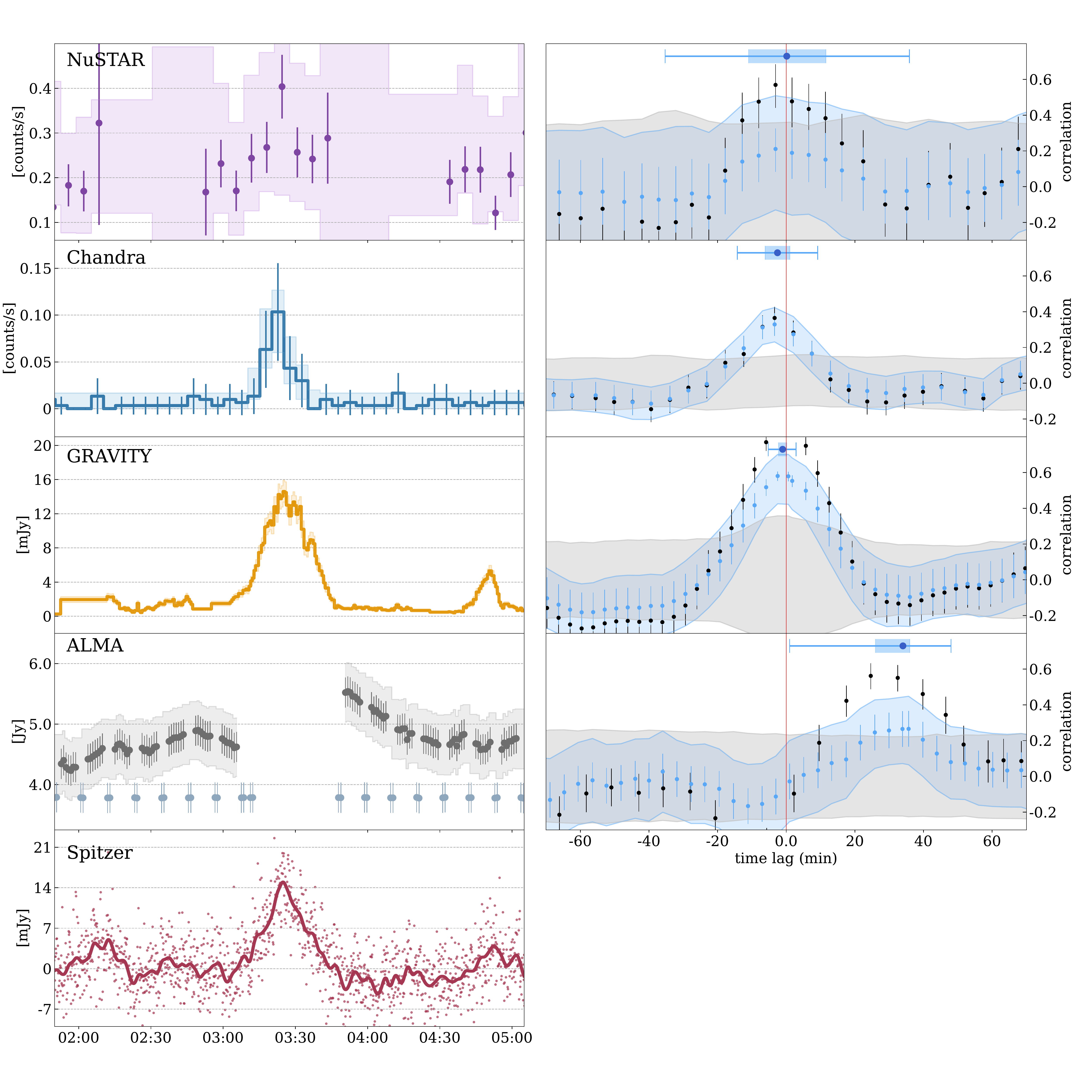}
\caption{Results from running \texttt{ZDCF} on the NuSTAR (purple), Chandra (blue), GRAVITY (orange), and ALMA (grey) light curves against the respective Spitzer (red) light curve on July 17/18, 2019. \textit{Left panels}: Regions of the multiwavelength light curves during the X-ray/NIR flare. Their respective envelopes show the 95\% range of the \iterations Monte Carlo (MC) realizations. The x-axis displays the UTC time since the start of the Spitzer/IRAC observations. The bottom panels show the \texttt{ZDCF}. \textit{Left panels}: The blue points are the average cross correlation of all \iterations MC realizations with the blue envelope capturing the 95\% range. The grey envelope is the 95\% range from the IR MC realizations with \iterations realizations of simulated noise consistent with the characteristics of the second light curve's emission (no flares). The significant time lags and confidence intervals are plotted as a single blue point in each panel, with the 68\% interval represented by the blue shaded box, and the 99.7\% interval represented with the thin error bar.}
\label{fig:zdcf_all}
\end{figure*}

\subsection{Multi-wavelength Timing}\label{CrossCorrelation}

During the Chandra X-ray flare on July 18, the emission from Sgr A* at 4.5 $\mu$m  and 2.2 $\mu$m rose within minutes of the X-ray peak. Nearly simultaneously, NuSTAR detected moderate X-ray variability through a measurement of increased count rate in a single 6 minute bin. At 340 GHz, ALMA observations also captured part of this flare, but missed the peak (Figure \ref{fig:zdcf_all}). 

\par To quantify lags between the peaks of potentially associated activity in the these observations we followed \cite{2019ApJ...871..161B}. We utilized the {\sc Fortran 95} implementation\footnote{Found at: www.weizmann.ac.il/weizsites/tal/research/software/} of the z-transform discrete correlation function \citep[\texttt{ZDCF};][]{1997ASSL..218..163A}. This tool estimates the cross-correlation function of two inputs without penalty for having a sparse or unevenly sampled light curve. We cross-correlated all observations relative to the simultaneous 4.5 $\mu$m Spitzer light curves binned at 3.5 min (red in Figures \ref{fig:lightcurves} and \ref{fig:zdcf_all}), which cover nearly all of the observing time of the other observatories. 

\begin{figure*}
\centering
\includegraphics[width=0.9\textwidth]{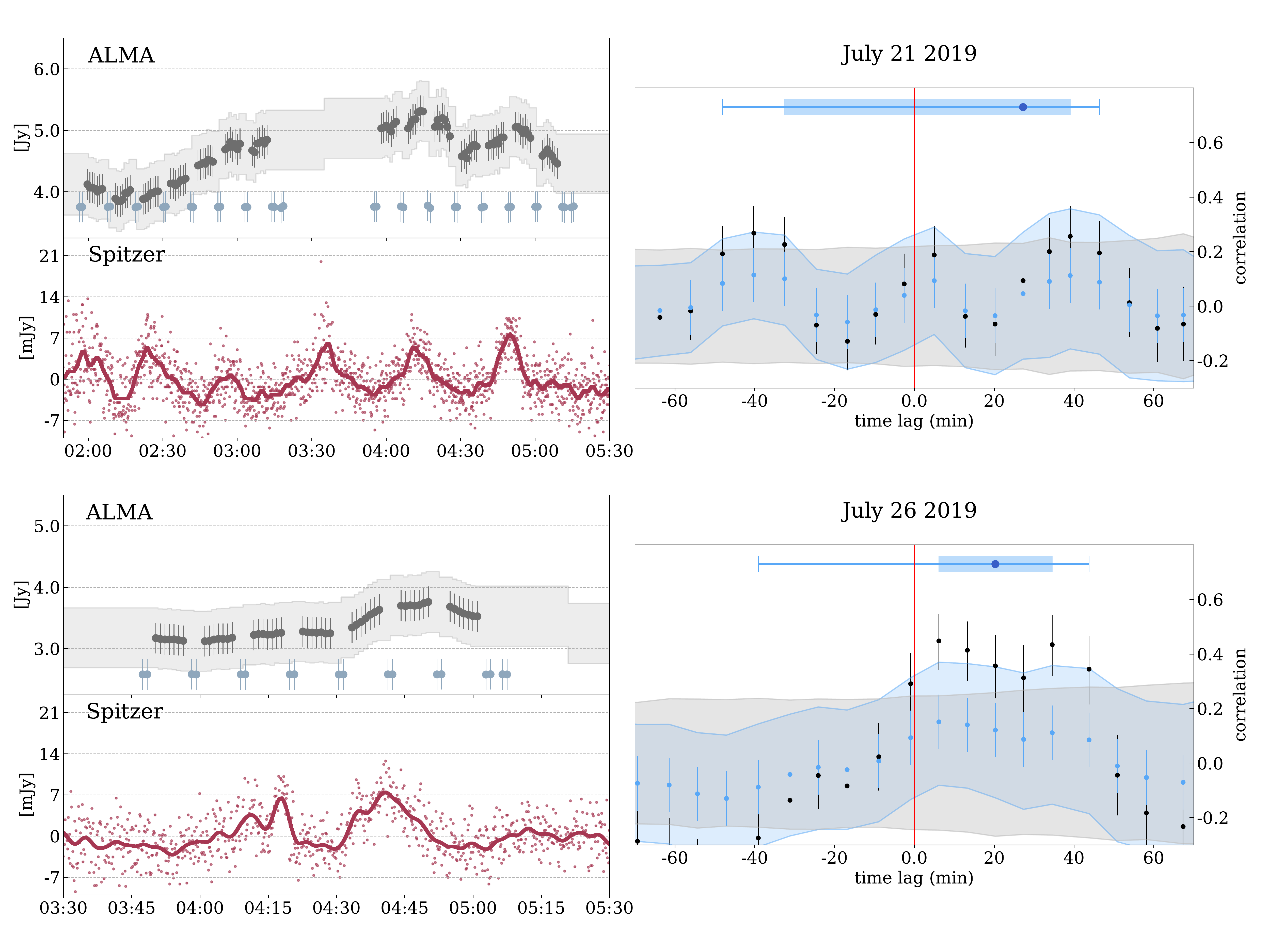}
\caption{Results from running \texttt{ZDCF} on the ALMA (grey) light curves against the respective Spitzer (red) light curve on July 21 and 26, 2019. \textit{Left panels}: Regions of the light curves portions where we see significant IR activity in the overlapping data. The grey envelopes show the 95\% range of the \iterations Monte Carlo (MC) realizations. The x-axes display the UTC time since the start of the Spitzer/IRAC observations. \textit{Right panels}: The blue points are the average cross correlation of all \iterations MC realizations with the blue envelope capturing the 95\% range. The grey envelope is the 95\% range from the IR MC realizations with \iterations realizations of simulated noise consistent with the RMS of the ALMA light curve's emission (no flares). The significant time lags and confidence intervals are plotted as a single blue point in each panel, with the 68\% interval represented by the blue shaded box, and the 99.7\% interval represented with the thin error bar.}
\label{fig:zdcf_21_26}
\end{figure*}

\par To estimate the uncertainties in the measured time-lags, we cross-correlated each pair of data over \iterations Monte-Carlo iterations. Bins of 3.5 min were chosen for the Spitzer data to increase efficiency of the cross-correlation Monte-Carlo analysis. Experiments with smaller bins yielded time-lags consistent with the results presented here. The uncertainty on the time lags was determined from the distribution of the \iterations \texttt{ZDCF} peaks \citep[see \S 3.2 of][]{2019ApJ...871..161B}. The observed correlation function (black) displays a stronger signal of correlation than the spread of simulations (blue) because of the way the data points in the simulated light curves are chosen. Each data point in a simulated light curve is randomly selected from a Gaussian distribution centered on the observed flux value in that bin with a standard deviation equal to the 1-sigma errors on the measured data. Therefore, real correlations in the detailed shape of the light curve (e.g. a monotonic rise) may not be reproduced strongly in a given simulated instance. The height of the shaded blue regions above the simulated noise can therefore be seen as a pessimistic indicator of how real the correlation is. Thus the width of the distribution of peak locations drawn from the simulations can conservatively estimate the uncertainty on the time lag. A positive time lag corresponds to variability in the the \textit{NuSTAR, Chandra, GRAVITY,} or ALMA data \textit{lagging} the 4.5 $\mu$m Spitzer variability, while a negative time lag corresponds to variability \textit{leading} 4.5 $\mu$m.

\begin{figure*}
\centering
\includegraphics[width=0.85\textwidth]{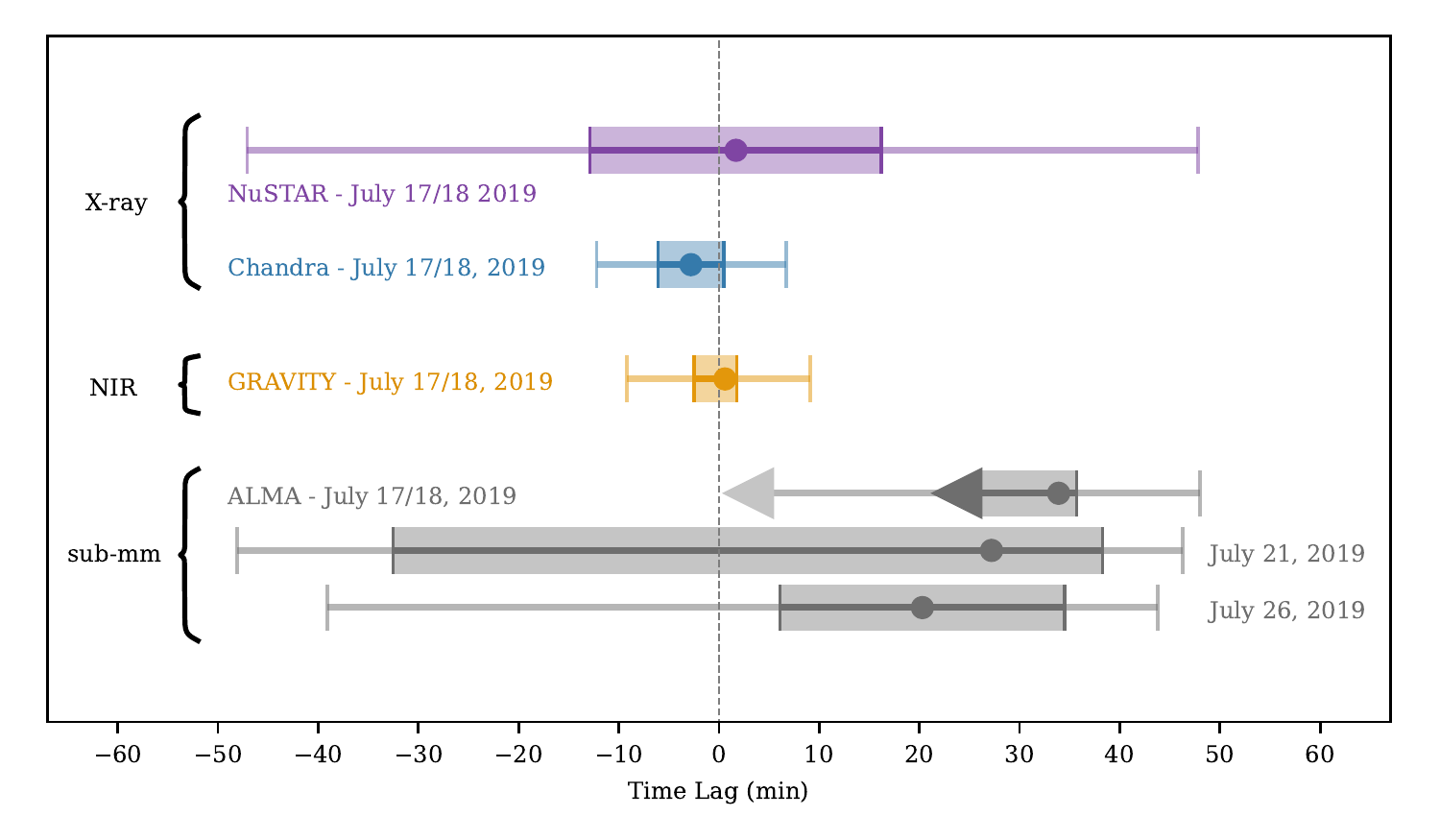}
\caption{Time lags between all multiwavelength observations and Spitzer 4.5$\mu$m light curves for the Sgr A* July 2019 campaign. The purple, blue, orange and grey points show the NuSTAR 3--70 keV, Chandra 2--8 keV, GRAVITY K-band (2.2 $\mu$m), and ALMA 340 GHz lags, respectively. The 68\% confidence intervals are represented by the shaded boxes, and the 99.7\% intervals are represented by the thin error bars. Because the measured submm lag on July 18 is an upper limit, the peak of the flare was not captured. }
\label{fig:timeLag}
\end{figure*}

\textit{Spitzer$-$NuSTAR}: Figure \ref{fig:zdcf_all} shows the results of running the \texttt{ZDCF} on the 2019 July 18 epoch of the Spitzer data and the 6min binned NuSTAR light curve. The measured time lag for the flare plotted in Figure \ref{fig:timeLag} and reported in Table \ref{tab:timeLags} at $+2\substack{+15\\ -15}$ minutes, is consistent with simultaneity but less significant than the Chandra X-ray measurement due to lower signal-to-noise and sensitivity in the data.

\textit{Spitzer$-$Chandra}: The second row of Figure \ref{fig:zdcf_all} shows the results of running the \texttt{ZDCF} on the 2019 July 17/18 epoch of the Spitzer data and 300s binned Chandra light curve. The measured time lag for the flare plotted in Figure \ref{fig:timeLag} and reported in Table \ref{tab:timeLags} at $-3\substack{+3 \\ -3}$ minutes is consistent with simultaneity.\footnote{An updated barycenter correction was applied to all reductions of the current and previous the Chandra data. This slightly altered the original results from \cite{2019ApJ...871..161B} but remained within the 1$\sigma$ uncertainties. The time lags for those NIR/X-ray epochs were recalculated and reported in \cite{Boyce_err} as well as here in Table \ref{tab:timeLags} and Figure \ref{fig:xray_ir_timeLag} in appendix \ref{appendix}.}

\textit{Spitzer$-$GRAVITY}: The results of running the \texttt{ZDCF} on the 2019 July 17/18 epoch of the Spitzer data against the 40s binned K-band GRAVITY light curve are also shown in Figure \ref{fig:zdcf_all}. The measured time lag for the flare is plotted in yellow in Figure \ref{fig:timeLag} and reported in Table \ref{tab:timeLags} at $+0\substack{+1\\ -3}$ minutes, consistent with simultaneity.

\textit{Spitzer$-$ALMA}: Figure \ref{fig:zdcf_all} shows the results of running the \texttt{ZDCF} on the 2019 July 17/18 epoch of the Spitzer and 340 GHz ALMA light curves, while the cross correlation of 2019 July 21$^{st}$, and 26$^{th}$ data sets are plotted in Figure \ref{fig:zdcf_21_26}. The measured time lags for the variabilty on each date are plotted in grey in Figure \ref{fig:timeLag} and reported in Table \ref{tab:timeLags}. 

\par Only data from July 18 show a X-ray flare with significant simultaneous NIR activity. During this window of 02:30$\sim$05:00, ALMA measured significant variability but missed the crucial window of 03:00$\sim$03:50 in which the NIR and X-ray flares occurred. The observed peak submm flux occured around 04:00 $-$ right after the window of missing data but at a time when NIR and X-ray flux levels had returned to typical quiescent rates. The result from the \texttt{ZDCF} on the July 18 Spitzer-vs-ALMA data is a measured time lag of $+34\substack{+2\\ -8}$ minutes at 68$\%$ confidence and $+34\substack{+14\\ -33}$ at 99$\%$ confidence. It is therefore likely that the peak of the submm flux lagged the NIR and X-ray variability by 10s of minutes, though we must interpret $\sim$35 min as an upper limit on the time-lag since the true peak was not observed.

\begin{table}[h!]
	\caption{Time delays with respect to 4.5 $\mu$m (Spitzer) for X-ray (NuSTAR, Chandra), $\sim$2 $\mu$m (GRAVITY), and 340 GHz (ALMA) variability.}
	\label{tab:timeLags}
\begin{tabularx}{0.475\textwidth}{@{\extracolsep{\fill}}llrr}
	
	\toprule
	Instrument & time lag (min) & 68\% interval & 99.7\% interval \\
	\toprule
	July 18 2019: \hspace{-7mm} &&&\\
	\midrule
	NuSTAR & $+2\substack{+15\\ -15}$ & ($-13$,$+16$) & ($-47$, $48$) \\
	Chandra & $-3\substack{+3 \\ -3}$ & ($-6$,$+0$) & ($-12$, $+7$) \\
	GRAVITY & $+0\substack{+1\\ -3}$ & ($-3$,$+1$) & ($-9$, $+9$) \\
	ALMA & $+34\substack{+2\\ -8}$ & ($+26$,$+36$) & ($+1$, $+48$) \\
	\midrule
	July 21 2019: &&&\\
	\midrule
	ALMA & $+27\substack{+12\\ -60}$ & ($-33$,$+39$) & ($-48$, $+46$) \\
	\midrule
	July 26 2019: &&&\\
	\midrule
	ALMA & $+20\substack{+14\\ -14}$ & ($+6$,$+35$) & ($-39$, $+44$) \\
	\bottomrule
	
\end{tabularx}

\textit{Note}: Positive values mean peaks lag Spitzer peaks.  Uncertainties on the time lag in the first column span the 68\% confidence interval on the \iterations MC runs. The second column displays the boundaries of this 68\% confidence interval, while the third column contains the 99.7\% confidence interval. 

\end{table}

\par Cross-correlating the Spitzer and ALMA light curves on July 21 and 26 followed the same method, and the results are displayed in Figure \ref{fig:zdcf_21_26}. Though there was not a significant X-ray flare, the NIR and submm show distinguishable variability. The cross-correlation of July 21 results in a lag of $+27\substack{+12\\ -58}$ minutes at 68$\%$ significance; a broad range that reaches over two marginally significant correlation peaks at around $-40$ and $+40$ minutes. The cross-correlation of July 26 results in a lag of  $+20\pm14$ minutes at 68$\%$ significance; consistent with the lag detected on July 18, but is also consistent with simultaneity $\sim$20$\%$ of the time. Figure \ref{fig:timeLag} summarizes the results.

\section{Discussion} \label{Discussion}

Variability in the NIR has been successfully described by the intermittent acceleration of electrons in a turbulent accretion flow, most often modelled as non-thermal synchrotron emission with a varying cooling cutoff. This is supported by the observed linear polarization of the IR emission  \citep{2006A&A...455....1E, 2006A&A...460...15M, 2007A&A...473..707M, 2007MNRAS.375..764T, 2007ApJ...668L..47Y, 2008A&A...479..625E, 2011A&A...525A.130W, 2015A&A...576A..20S}, the spectral index at high flux densities \citep[$\alpha \approx -0.6$;][]{2007ApJ...667..900H, 2011A&A...532A..26B, 2014IAUS..303..274W}, and the timescale of the variability, with factors of $\gtrsim$10 changes within $\sim$10 minutes \citep[e.g.,][]{2003Natur.425..934G, 2004ApJ...601L.159G,2018ApJ...863...15W}. 

The physical parameters of this turbulent acceleration of electrons (e.g., background magnetic field strength $B$, the Lorentz factor of the electrons $\gamma$, and the electron density n$_{e}$) and the details of the radiative processes linking the NIR variability to the X-ray flares are still uncertain. The processes often invoked to make this connection include (1) pure synchrotron from a sudden acceleration of electrons to a non-thermal distribution \citep[e.g.,][]{2001A&A...379L..13M,2009ApJ...698..676D, 2014ApJ...786...46B, 2017MNRAS.468.2447P}, (2) synchrotron self-Compton through the scattering of these non-thermal synchrotron photons up to X-ray energies \citep{2001A&A...379L..13M,2008A&A...479..625E,2012A&A...537A..52E,witzel+2021}, and (3) inverse Compton scattering of radio and submm photons from the synchrotron source produced by the persistent large population of thermal electrons \citep{2012AJ....144....1Y}. All these scenarios can include changes in the source's magnetic field ($B$), electron density (n$_{e}$), and Lorentz factors ($\gamma$). The most likely scenario may be some combination of multiple processes, but the unpredictable nature of flares from around accreting BHs limits data collection, and often the best way forward is testing one scenario at a time. 
\par More broadly, Sgr A*'s average SED is described by several varying components that could originate from different zones in the accretion flow. Though the connection between the NIR and X-ray is clear, it remains an open question whether submm variability could originate from the same source as the higher energy variability. Periods of increased submm variability can be described by separate, uncorrelated events that are occasionally coincident with NIR/X-ray flares. We ask whether the submm, NIR, and X-ray variability on July 18 could be explained through a single acceleration event, i.e. a single-zone modelled at the peak of the NIR/X-ray flare and tens of minutes later, when submm flux is observed to be declining from an unknown peak value.

To tackle this question, we re-examine three different scenarios of (1) and (2), in light of the total dataset from the campaign presented here, wherein (A: 0-SYNC-SYNC) non-thermal emission originating from a single source of accelerated electrons is responsible for the NIR and X-ray while contribution to the submm is negligible, (B: SYNC-SYNC-SSC) non-thermal synchrotron emission is responsible for the submm and NIR while the X-rays are produced through SSC processes, and (C: SYNC-SSC-SSC) submm flux density is due to a non-thermal population of electrons emitting synchrotron radiation while both the NIR and X-ray are dominated by the SSC emission. IC scattering of external thermal submm photons (3) is not examined. All SEDs discussed in the following sections are produced with \texttt{flaremodel} \citep{Dallilar+2022}, a code for numerically modelling one-zone synchrotron sources\footnote{Available at \url{https://github.com/ydallilar/flaremodel}}. 

Our multi-wavelength time-resolved data constrain the evolution of the source as these electrons cool and/or are continuously accelerated. We are motivated to test these single-zone descriptions because they do not over-fit our data by introducing complex geometries and because flaring in the NIR has been successfully described as originating from a compact, orbiting hot-spot on horizons scales \citep{2020A&A...635A.143G}. Once electrons are accelerated, they may cool via several channels that would affect the accretion structure around a BH \citep[e.g., synchrotron, bremsstrahlung, and inverse Compton processes,][]{2020MNRAS.499.3178Y}. Here we examine one possibility via cooling under adiabatic expansion, in which a uniform and spherical cloud of relativistic electrons is expanding and the cooling applies to electrons of all energies at the same rate set by the expansion speed. We refer to the time of the NIR/X-ray peak as t$=$0, and the time of the measured 340 GHz ``peak" as t$=$35 min.

\begin{figure}
\centering
\includegraphics[width=0.475\textwidth]{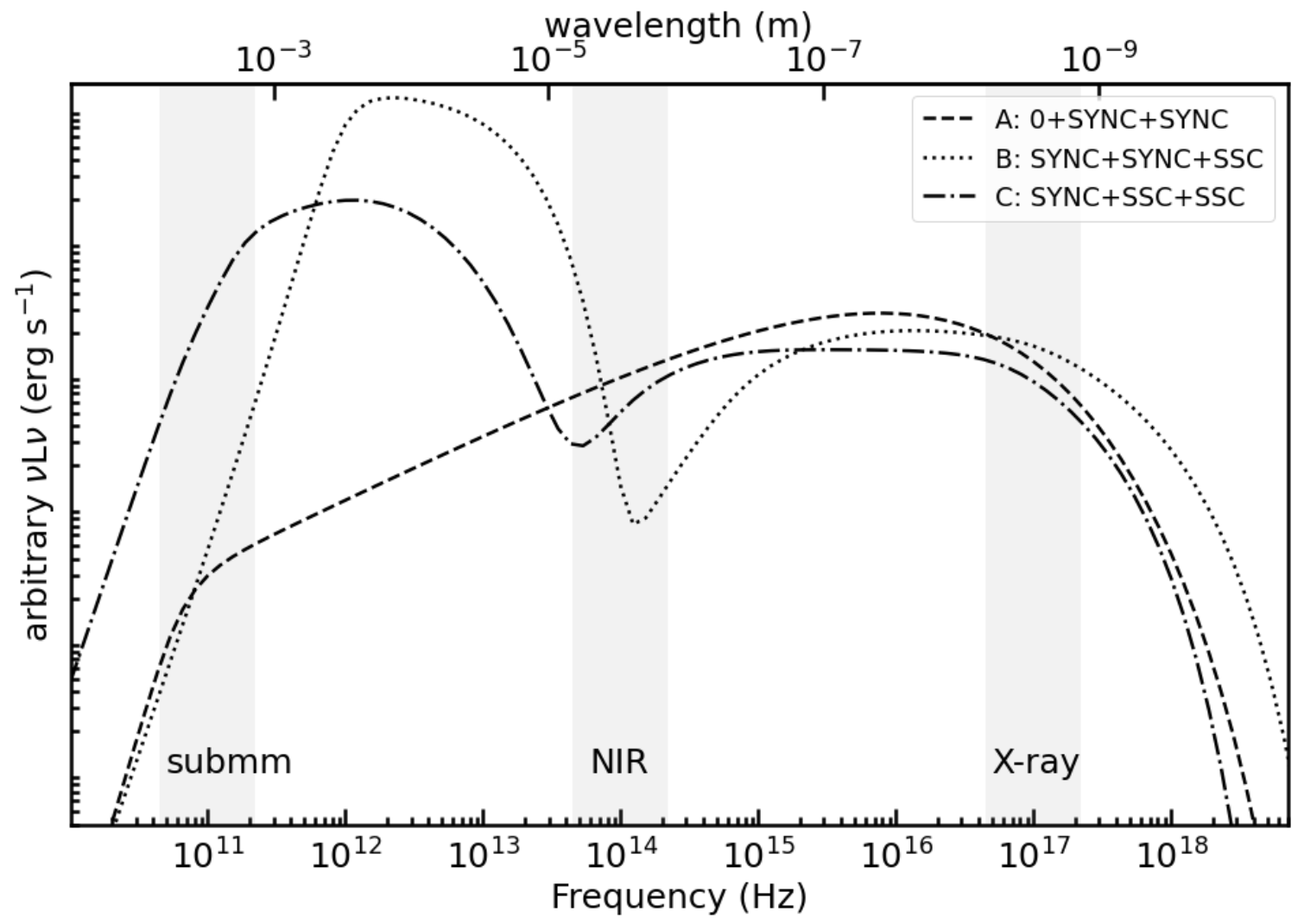}
\caption{Illustration of the three example SED models at the time of the NIR/X-ray peak (t$=$0). (A): The dashed line represents model 0-SYNC-SYNC in which both the NIR and X-ray flux is described by a synchrotron source that contributes negligibly to the submm. (B): The dotted line represents model SYNC-SYNC-SSC in which the optically thick cutoff of the non-thermal SYNC component contributes to the submm, the varying optically thin cutoff of the same SYNC component contributes to the NIR, and the X-ray variability is produced through SSC. (C): The dash-dotted line represents model SYNC-SSC-SSC, in which submm flux can be explained through the optically thick SYNC, NIR flux is dominated by SSC, and the X-rays are also produced by SSC. The models illustrate the shape of the SEDs, but the relative vertical position (flux) of these example curves is arbitrary. Details described \cite{Dallilar+2022}.}
\label{fig:models}
\end{figure}

\subsection{(A) +SYNC+SYNC: An evolving synchrotron source}

We consider the scenario where both the NIR and X-ray are produced by a single synchrotron spectrum originating from particle acceleration events involving magnetic re-connection and shocks in the accretion flow \citep[e.g.,][]{2001A&A...379L..13M,2009ApJ...698..676D, 2014ApJ...786...46B, 2017MNRAS.468.2447P, 2017ApJ...850...29R}. An example of this scenario is plotted as the dashed line in Figure \ref{fig:models}. With both bands being produced by a single non-thermal synchrotron source, the synchrotron cooling time in the NIR would far exceed the X-rays and the source would require sustained particle acceleration to produce observed X-ray flare durations of up to $\sim$1 hour. In this scenario, rapid synchrotron cooling will cause fading in the higher energy X-rays sooner than in the NIR \citep[e.g., see \S 4.1 of ][]{2010ApJ...725..450D}. This could manifest as a simultaneous rise with a time-delay between the X-ray and NIR flare ``centres" of a few to 10s of minutes if the time resolution and signal-to-noise of of our observed X-ray light curves were high enough \citep{2010ApJ...725..450D}.

Cooling the best-fit synchrotron model of \cite{seb_flare_paper} at time t$=$0 (via any cooling process) would result in a decrease in flux across the SED and would not produce appreciable flux in the submm. Therefore, if the NIR and X-ray variability is due to a purely SYNC component \citep[see, e.g.,][]{2017MNRAS.468.2447P} then that same source could not explain the observed $\sim2$ Jy increase in flux density at 340 GHz. The variability at these wavelength regimes must be physically uncorrelated or involve more complex models containing multiple zones of accelerated electrons in complex geometries. On the other hand, more complex models or geometries are difficult to include in the scenario wherein the submm flux correlates with NIR flares originating from a compact orbiting hot-spot on horizon scales \citep{2020A&A...635A.143G}.

In summary, the best-fit cooled SYNC model described in \cite{seb_flare_paper} accounts for the X-ray and NIR variability and does not require unusually large electron densities. However, as this synchrotron source cools flux at all wavelengths decreases. A simultaneous or delayed $1\sim2$ Jy increase in the submm flux density requires invoking multiple non-thermal populations of accelerated electrons and would not be physically correlated through the evolution of the same SYNC source responsible for the NIR and X-rays.

\subsection{(B) SYNC+SYNC+SSC: An adiabatically cooling synchrotron source}

\cite{witzel+2021} considers a simple physical model of a compact synchrotron component in Sgr A*'s accretion flow undergoing a sequence of: 
\begin{enumerate}
    \item injection of non-thermal electrons giving rise to detectable submm and NIR emission,
    \item further injection, compression of the source, and increasing magnetic flux resulting in higher NIR levels and detectable X-ray emission, and
    \item adiabatic expansion with little to no injection giving rise to maximum submm emission. \citep[For a deeper description, see Section 4 of][along with their Figure 11.]{witzel+2021}
\end{enumerate}

This sequence is based on the scenario that there exists a variable synchrotron spectrum arising from populations of non-thermal (accelerated) electrons in addition to the dominant thermal synchrotron radio component of Sgr A*’s SED. The NIR variability is then primarily due to the rapidly varying cooling cutoff of this spectrum. Correlated X-ray variability arises from the resulting SSC spectrum (with the high temporal frequency variability suppressed). This “slow” variability in the SSC X-rays is therefore related to physical changes in the synchrotron source itself (i.e., source size $\theta$, magnetic flux $B$, and self-absorption properties which manifest in changes to the location of the peak flux and self-absorption frequency turnover of the synchrotron spectrum at submm wavelengths). Delayed submm variability relative to NIR/X-ray is attributed to these physical changes in the source (e.g. cooling causes the SYNC component to shift to longer wavelengths).

This model predicts a delay in peak submm flux density on the order of $20-30$ minutes, consistent with our upper limit of $\sim$35min. It also describes the correlation of the majority of NIR and X-ray flares in the literature. \cite{Michail+2021} consider an analogous description of the synchrotron source for the case that the 2018 July 18 NIR and submm emission were simultaneous and find that conditions with p$=2.5$ describes the submm/IR flux increase well. In this case, simultaneity in submm and NIR could occur if conditions in the accretion flow produced a SYNC source with “optically thin” emission reaching from the submm regime to the NIR. This is incompatible with our observations in two ways:
First, the SYNC spectrum whose peak is near 340 GHz and broadly reaches the NIR does not produce SSC in the correct regime to fit the NIR/X-ray data. 
Second, this SYNC spectrum rising in the submm and reaching the NIR would not have the spectral index observed in the IR.

For typical ranges of physical parameters most of the variable NIR flux is produced by the optically thin cutoff of the synchrotron component and is described by relatively steep flux spectral index ($F_{\nu}\propto\nu^{\alpha}$) in the range $-2.0\lesssim\alpha\lesssim-0.8$, resulting in a negative or flat luminosity spectral index ($\beta=\alpha+1$). An example of this model (with a steep negative spectral index) is plotted as the dotted line in Figure \ref{fig:models}. 

\begin{figure*}
\centering
\includegraphics[width=0.7\textwidth]{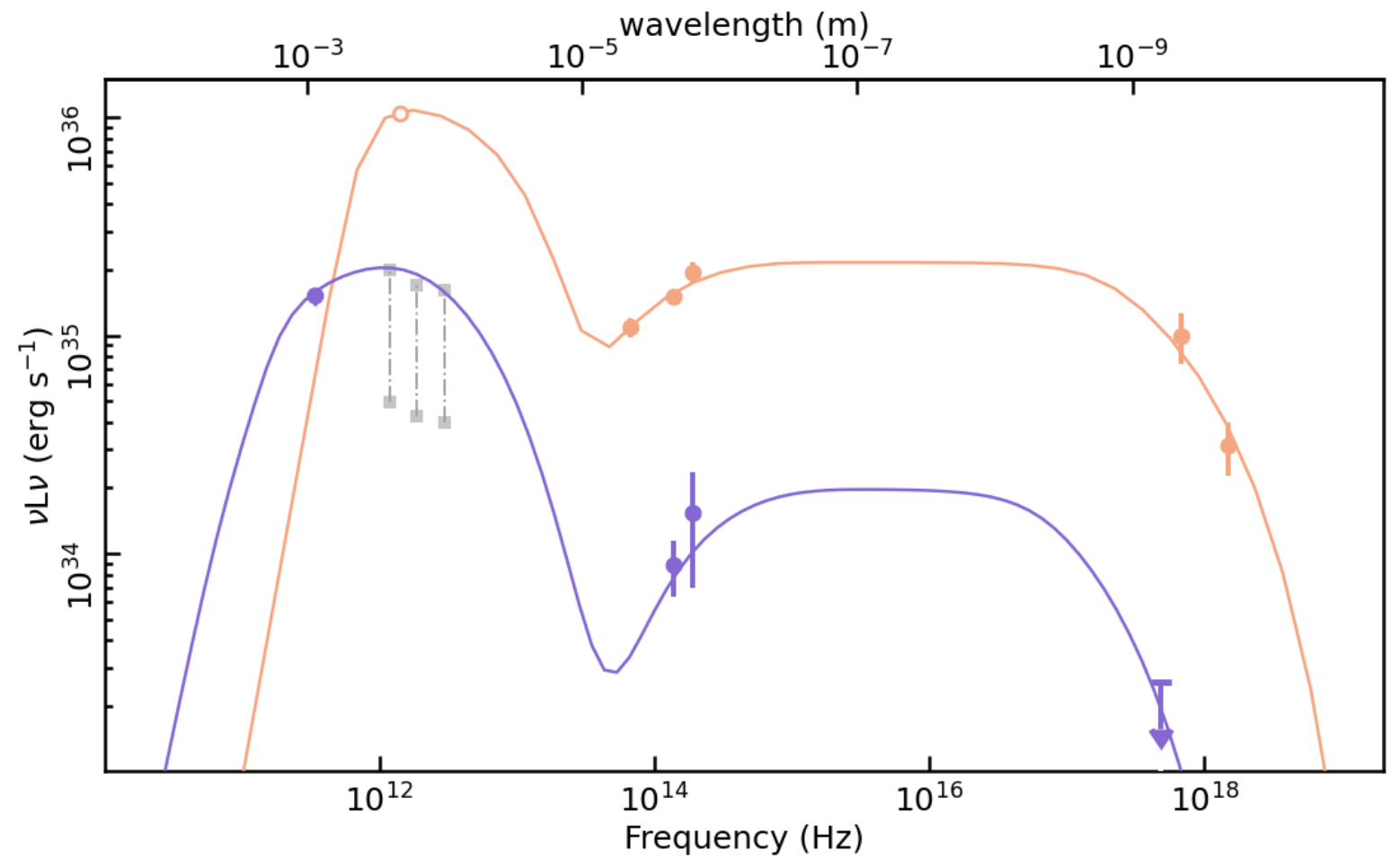}
\caption{Snapshots of the time evolution of scenario (C) SYNC-SSC-SSC, fit under adiabatic expansion. Solid orange points are measured values at the peak of the NIR and X-ray flare, solid purple points are measured at the presumed ``peak" of the 340 GHz flux $\sim$35 min later. The open orange point at 1400 GHz is used as the starting point in the adiabatic expansion calculation described below. The historic quiescent SED in radio/submm is plotted in light grey with a thermal synchrotron component fit to these data as the grey dashed line. The solid lines are the best-fit models with the thermal component included. Parameters for the fits are reported in Table \ref{tab:SSC-fit}. Observed values are tabulated in Table \ref{tab:sed}.}
\label{fig:sed}
\end{figure*}

\cite{seb_flare_paper} measured the evolution of the NIR spectral index of the July 18 flare and found that the GRAVITY K$-$band to Spitzer M$-$band slope varied in the range $\alpha_{K-M}=$ [$-$0.8, 0.0], consistent with the canonical NIR spectral index for bright ﬂares of $\alpha_{\mathrm{NIR}}\sim-$0.65 \citep[i.e., luminosity rising with shorter wavelengths]{2005ApJ...628..246E, 2005ApJ...635.1087G, Gillessen_2006, 2006ApJ...642L.145K, 2007ApJ...667..900H, 2011A&A...532A..26B, 2014IAUS..303..274W}. This is reflected in Figure \ref{fig:sed}, where the orange points in the NIR band have positive $\beta$. Since this flare has a NIR spectral index $\alpha\sim-0.6$ at its peak (Figure \ref{fig:sed}), we favor descriptions with positive luminosity photon indices ($\nu L_{\nu}\propto\nu^{\beta}$, $\beta=\alpha+1$).

In summary, though this scenario could explain the temporal evolution of the correlated submm, NIR, and X-ray flux densities, the spectral index in the NIR disfavours a scenario in which the NIR is dominated by the optically thin component of the SYNC spectrum.

\subsection{(C) SYNC+SSC+SSC: An adiabatically cooling synchrotron source}

Another possibility is that both the X-ray and NIR flux may be dominated by SSC flux (i.e., photons being scattered to higher energies through interaction with the electrons producing the non-thermal synchrotron in the submm). In this scenario the NIR flux would derive from the rising side of the SSC component, rather than the optically-thin edge of the SYNC component (which is now shifted toward even longer wavelengths). An example of this SED is illustrated as the dashed-dotted line ``C" in Figure \ref{fig:models}. 

Since the SYNC+SSC+SSC scenario predicts the correct range of NIR spectral indices, we fit this model with a synchrotron source that produces the 340 GHz flux increase that can evolve under adiabatic expansion. If adiabatic cooling is dominant, the SYNC source expands and cools (without further electron injection), causing the turn-over of the SYNC component to march down to lower frequency as it fades. This results in a delay at longer wavelengths. If the true peak of the submm rise was simultaneous with the NIR/X-ray, the SYNC component of the SED must have peaked near 340 GHz. Such a SYNC spectrum could not then produce bright enough SSC emission to match the NIR/X-ray observations. We therefore consider the scenario in which the peak of the submm emission was delayed by 10s of minutes.

To test this scenario and leverage the submm flux measured with a delay, we use the methodology first described in \cite{1966Natur.211.1131V} to parameterize the behaviour of the peak of the non-thermal SYNC component under adiabatic cooling. This method has been applied to interpret Sgr A* variability in the past \citep[e.g.][]{2006ApJ...650..189Y,2008A&A...492..337E}

The flux density as a function of frequency ($\nu$) is parameterized as:
\begin{equation}\label{eq:flux}
    S(\nu,\rho)=(\nu/\nu_{m})^{5/2}\rho^{3}\frac{\left[1-\exp\Big\{-\tau_{m}\left(\frac{\nu}{\nu_{m}}\right)^{-(p+4)/2}\rho^{-(2p+3)}\Big\}\right]}{\left[1-\exp\left(-\tau_{m}\right)\right]}
\end{equation}

where $\nu_{m}$ is the frequency at which the flux density maximum of the spectrum occurs, $p$ is the slope of the electron distribution, $\tau_{m}$ is the optical depth corresponding to the frequency at which the flux density is maximum and $\rho$ is the relative radius of the source, which can be parameterized in terms of the expansion velocity $v_{\mathrm{exp}}$, time ($t$), initial source size $R_{0}$, and a deceleration parameter $\beta$ (kept at standard value of 1.0 in our analyses):

\begin{equation}\label{eq:v}
    \rho=\left(\frac{1+v_{\mathrm{exp}}c(t-t_{0})}{R_{0}\beta}\right)^{\beta}
\end{equation}

To describe the broad-band SED, we numerically implement the SYNC-SSC model described in \cite{Dallilar+2022}, based on a non-thermal power-law-distributed electron energy distribution. The physical parameters of this single-zone non-thermal synchrotron model are the electron density (n$_{e}\times1$cm$^{3}$, the projected radius ($R$, $\mu$as), the magnetic field ($B$, G), the power law slope of the electron distribution ($p$), the maximum Lorentz factor ($\gamma_{\mathrm{max}}$), and the minimum Lorentz factor ($\gamma_{\mathrm{min}}$).

Plotted in orange points in Figure \ref{fig:sed} are the observed X-ray and NIR data at the time of their peak ($t=0$). Due to the gap in the observing window, we do not have a simultaneous measurement at 340 GHz. However, under the assumption that there is a significant time lag of $\sim$35 min or less in the peak of the submm flux, the 340 GHz flux at $t=0$ must be fainter than $\sim$2 Jy (excess flux above historic quiescence; 5.5$\times10^{34}$ erg s$^{-1}$ at 340 GHz). The orange line is the best-fit SSC-SSC SED that satisfies this constraint with $\chi_{\mathrm{red}}^{2}$ of 2.3. Experimenting with the errors on the data, we find that the high H-band measurement prevents the fit from reaching $\chi_{\mathrm{red}}^{2}\sim1$. Doubling the uncertainty on this point would result in $\chi_{\mathrm{red}}^{2}=1.1$ with very similar values to those listed in Table \ref{tab:SSC-fit}.

In purple are the constraints in the NIR/X-ray once their flux has faded (at ~35 min past peak) as well as the measured ``peak" flux at 340 GHz. Fitting these data with the SYNC+SSC+SSC SED + SYNC thermal component (grey) yields  $\chi_{\mathrm{red}}^{2}$ 0.5, Physical parameters of these best-fits are tabulated in Table \ref{tab:SSC-fit}.

\begin{table}[h!]
	\caption{Best fit parameters of scenario C: SYNC+SSC+SSC.}
	\label{tab:SSC-fit}
\begin{tabularx}{0.475\textwidth}{@{\extracolsep{\fill}}lcc}
	
	\toprule
	 & $t=0$ & $t=35$ min \\
	\toprule
	$\log$ (n$_{e} \times 1 $cm$^{-3}$)& $10.1\pm0.8$ & $9\dagger$ \\
	$R$ ($\mu$as)* & $11.2\pm2.1$ & $21\pm2$ \\
	$B$ (G) & $25\pm44$ & $3.1\pm0.8$ \\
	$p$ & $3\dagger$ & $3\dagger$ \\
	$\gamma_{\mathrm{max}}$ & $320\pm110$ & $410\pm130$ \\
	$\gamma_{\mathrm{min}}$ & $3.8\pm1.3$ & $10.8\pm3.4$ \\
	\midrule
	$\chi_{\mathrm{red}}^{2}$ & $2.3$ & $0.5$\\
	\bottomrule
	
\end{tabularx}
$*$: 1 $\mu$as $=$ 0.0082 AU\\
$\dagger$: value fixed

\end{table} 

\begin{figure}
\centering
\includegraphics[width=0.475\textwidth]{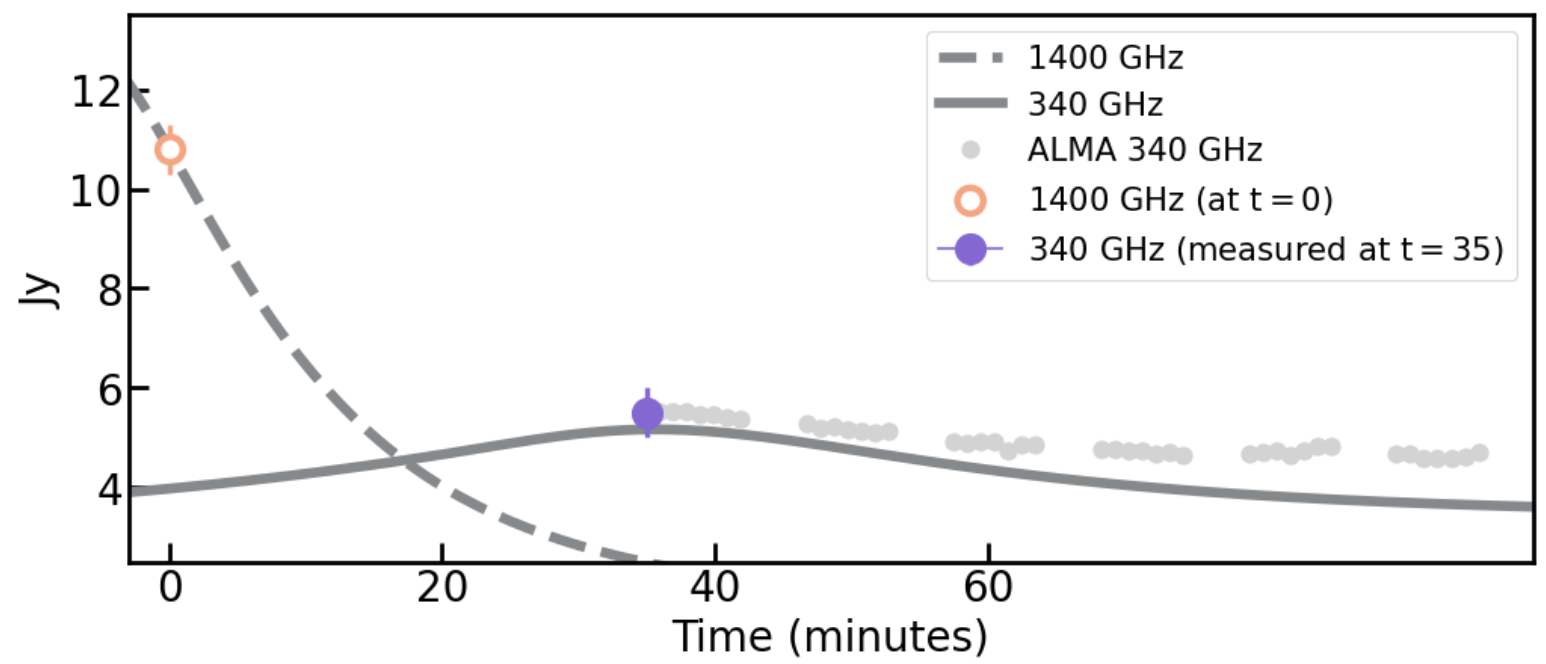}
\caption{Light curves derived from equations \ref{eq:flux}, and \ref{eq:v} offset with the constant flux values originating from the putative constant synchrotron component that arises from a thermal distribution of electrons (dashed grey model in Figure \ref{fig:sed}; 1.5 Jy at 1400 GHz and 3.4 Jy at 340 GHz). A value of 9.3 Jy at 1.4 THz (10.8 Jy from Figure \ref{fig:sed} $-$ 1.5 Jy from the thermal component) is used in the calculation and evolved forward in time with expansion velocity v$_{\mathrm{exp}}=0.014$c.  The purple data point (5.5 Jy) is the measured 340 GHz flux with the thermal synchrotron component (3.4 Jy) included.}
\label{fig:vdL_1}
\end{figure}

Taking the best-fit radius at $t=0$ ($R_{0}\sim1.1\times$R$_{\mathrm{S}}$) and the peak flux at 1400 GHz (10.8$-$1.5$=$9.3 Jy after subtracting the thermal component from the peak in Figure \ref{fig:sed}) we apply Equations \ref{eq:flux}, and \ref{eq:v} \citep{1966Natur.211.1131V} to match the peak flux in 340 GHz at $t=35$ min. With the initial size of the region, $R_{0}$, set at the best fit value, we can vary the expansion speed and find that a value of $v_{\mathrm{exp}}\sim$0.014c reproduces the flux observed at the peak in 340 GHz (see Fig \ref{fig:vdL_1}). This speed is consistent with other estimates of $v_{\mathrm{exp}}\sim$0.003c$-$0.02c found under the interpretation of an expanding plasmon \citep[in the cm]{2006ApJ...650..189Y} and \citep[NIR-submm]{2006A&A...450..535E,2008A&A...492..337E,2008ApJ...682..373M,2012A&A...537A..52E}. This calculation relies on the assumption that the peak in 340 GHz occurred at $t=35$ min. If the peak happened earlier, we would require an even faster expansion speed to match the measured flux.

Scenario ``C" (SYNC+SSC+SSC) can be interpreted as a particularly unusual version of scenario ``B", in which the same single zone model and radiation mechanisms could produce typical flux variations in the submm, NIR, and X-ray. In this picture the July 18 event's unusually high submm flux is explained through uniquely high electron densities and a prediction of bright emission in the THz regime.

This interpretation relies on the validity of two unique characteristics: First, there must have been very high flux at $\sim$THz frequencies during the flare, something that has not been reported in campaigns aiming to characterize the quiescent THz spectrum \citep{2018ApJ...862..129V,Bower+2019}, though at 850 GHz \cite{1997ApJ...490L..77S} report a measurement of $\sim3$ Jy ($2\times10^{35}$ erg s$^{-1}$). An updated study on the flux density distribution at submm-THz is required to determine the likelihood of observing such a flare based on past observations at these frequencies. Second, electron densities in the SYNC source must have been several orders of magnitude higher than the implied densities of the average accretion flow from radio polarization studies \citep[$\log{n}_{e}$ $\sim$ 10 compared to e.g. $\sim$ 7;][]{Bower+2019}, which could be easier to achieve if Sgr A* were in an unusual state of increased accretion. 

Average accretion rates for Sgr A* are estimated from the rotation measure in quiescent submm observations, the last of which was \cite{Bower+2018}, based on data obtained in 2016. The rotation measure has been observed to have short term variability and most estimates of Sgr A*'s accretion rate are cited from the value averaged over the long term ($\sim$years). Since then there have been hints that Sgr A*'s accretion state may not be so constant, particularly supported by the incredibly bright NIR flare observed in early 2019 \citep{2019ApJ...882L..27D}, which fell outside of all previously-parameterized flux density distributions. If Sgr A* was indeed in a state of elevated accretion in 2019, then this could explain how this event is distinct from most previously observed flares. That is, a high sub-THz flux may be more easily achievable if electron densities as a whole were increased, allowing flaring conditions with $\log{n}_{e}$ $\sim$10.

Finally, the assumption that we have captured the peak of the 340 GHz flare is a large one. The start of the observing window around 03:45 catches the light curve in a descending state, with no indication of a turn-over (Figure \ref{fig:lightcurves}). If we have not captured the peak of the flare, that would allow the SYNC component in this SED to extend to lower energies at the time of the NIR/X-ray peak, though it would still remain significantly higher than the previously observed flux levels at these frequencies.

Fitting the temporally-resolved SED over six time-steps in the NIR and X-ray, \cite{seb_flare_paper} conclude that the particle densities necessary for ``C" SYNC+SSC+SSC ($\gg10^{9}$ cm$^{-3}$) would be extremely unlikely given the typical, average electron densities derived from modeling the radio to submm SED of Sgr A* with synchrotron emission from a thermal electron distribution \citep[\textnormal{ambient} n$_{e}< 10^{7}$ cm$^{-3}$][]{Bower+2019}. To fit an SED like ``A" (0+SYNC+SYNC), a strong acceleration event is necessary ($\gamma_{\mathrm{max}}\gg10^{4}$), but the physical parameters of the source (including n$_{e}$) remain consistent with predictions from the literature.

Of course, adiabatic expansion is not the only scenario in which delayed and correlated emission between frequencies can arise. Interpreting 20–40 minute delays in the 20 to 40 GHz regime, \cite{2009A&A...496...77F} observed a frequency dependence of VLBI sizes and saw evidence for a relativistic outflow. Meanwhile \cite{2015A&A...576A..41B} derived relativistic outflow velocities of up to $\sim$0.77c, through the progression of variability maxima from 100 to 19 GHz, and interpret this as a jet. Finally, it is always possible that the submm variability is not physically correlated with the NIR and arises from another component altogether. For example, single-zone modelling of M87's jet and active galactic nucleus cannot fully describe the broadband SED, albeit the data most driving this conclusion are the high-energy $\gamma$-rays \citep{EHT_M87_MWL_2021}

\par In summary, this single-zone adiabatic expansion model fits the data, with the caveat that the inferred submm-THz flux at $t=0$ would have been much brighter than most observations at these frequencies. Accepting the plausibility of the scenario requires an electron density that would be extraordinarily high in comparison with estimated densities responsible for Sgr A*'s average accretion rate. 

\section{Summary}

We report observations from a multi-wavelength campaign that simultaneously observed Sgr A* in July 2019 from the submm, to NIR, to X-ray. Cross correlating the light curves against the Spitzer NIR light curve on each date, we report the measured time lags between each wavelength. 

\begin{itemize}
    \item On July 17$-$18, a moderately bright NIR flare captured by Spitzer (4.5$\mu$m) and GRAVITY (2.2$\mu$m) occurred simultaneously with a faint X-ray flare captured by NuSTAR ($3-70$keV) and Chandra ($2-8$keV). Overlapping coverage at 340 GHz from ALMA missed the peak of the submm flare, but reveals very bright correlated flux $\sim$35 min after the NIR/X-ray peak.
    \item On July 21, correlated submm/NIR flux variability remains consistent with simultaneity (no time lag).
    \item On July 26, we report a measured time lag of $\sim$20 min between correlated submm and NIR variability with 68$\%$ confidence, though consistent with simultaneity at 98\% confidence.
\end{itemize}

The flux and timing properties of the July 17$-$18 flare are considered in the context of three scenarios:  ``A" both NIR and X-ray due to emission from a synchrotron source, ``B" submm and NIR due to a synchrotron source while X-ray arises as synchrotron self-Compton emission, and ``C" submm due to a synchrotron source while both NIR/X-ray arise from synchrotron self-Compton. We are limited in what we constrain because we have not captured the peak of the 340 GHz flare, and can only measure an upper limit on the time-lag between it and the NIR. This event is particularly interesting because the submm flux is notably high ($\sim$5.5 Jy, very rarely observed at these frequencies), so if the peak is even higher, this could indicate that the radiative processes are non-typical when compared to conditions responsible for historic variability. In the scenario in which the submm and NIR/X-ray variability are not physically correlated, a SYNC source fitted to the NIR/X-ray (scenario ``A") is allowed and does not require extraordinarily large electron densities \citep[][]{seb_flare_paper}. 

To leverage the potentially delayed submm flux, we consider whether a synchrotron source cooled through adiabatic expansion can self-consistently describe the submm increase and the NIR/X-ray flux at peak and after. Consistent with our measurement, \cite{Michail+2021} report an upper limit on the time lag of less than 30 minutes. They also analyze the submm and mid-IR emission using adiabatically expanding synchrotron plasma models and find two cases that can describe the data. The first is a SYNC source with p $=$ 2.5 responsible for simultaneous rise in the submm and NIR (analogous to scenario ``B" SYNC+SYNC+SSC). We disfavour this scenario primarily since the predicted NIR spectral index is in tension with the observations, but also since a simultaneous rise in the submm and NIR would require a SYNC spectrum whose peak is near 340 GHz and broadly reaches the NIR, which does not produce SSC in the correct regime to fit the X-ray data. In their second case, a SYNC source with p $>$ 2.8 has optically thick plasma conditions that evolve to optically thin in the submm on the time scale of 10s of minutes (analogous to scenario ``C" SYNC+SSC+SSC). We find that this adiabatic expansion scenario producing SSC emission in the NIR and X-rays (scenario C) works only under the conditions that a very high submm/THz peak would occur at the time of the NIR/X-ray peak and that the electron density reaches $\log{n}_{e}$ $\sim$10. 

Narrowing down the radiation mechanism powering and connecting variability across wavelength regimes brings the field closer to accurately describing the physical mechanisms that power the dramatic flux changes originating near the event horizon. Simultaneous, multi-wavelength observations of Sgr A* at all accessible frequencies remain essential to differentiate between various radiation mechanisms. Such observational campaigns are key to comparing to the state-of-the-art general-relativistic magneto-hydrodynamic (GRMHD) simulations that can model details of accreting plasma in this extreme environment, where high resolution simulations have shown that sufficiently energetic plasma can be accelerated through magnetic reconnection \citep{2022ApJ...924L..32R}. In particular, continued coordination between submm-radio observatories and the NIR/X-ray will strengthen or rule out the interpretation that these variable signals are physically connected. If simultaneous observations at THz frequencies are also captured during submm/NIR/X-ray variability, one could definitively constrain models in which the cooling SYNC component is responsible for the submm flux density increase and is correlated with NIR/X-ray SSC emission. Finally, coordinated multi-wavelength campaigns with the EHT and VLTI/GRAVITY will be key to interpreting the increasingly detailed view of this accreting SMBH on horizon scales.

\acknowledgments

We thank the GRAVITY collaboration for sharing the flux data for July 17/18 \citep{seb_flare_paper} and their valuable feedback on the analysis and text. We are grateful to Yigit Dallilar for help in the utilization of the \texttt{flaremodel} SED code \citep{Dallilar+2022}, his expert guidance in the implementation of adiabatic expansion codes, and thank him for numerous fruitful scientific discussions. We are thankful for Gabriele Ponti's helpful input and insightful discussions. We thank Eduardo Ros for his careful reading of the text and valuable clarifying comments. The authors are grateful for access to the privileged location of the high-altitude plateau Chajnantor in the land of the indigenous Likanantai people on which the ALMA telescope sits. HB and DH acknowledge and thank the diverse indigenous people on whose land our home institutions reside in Tiohti:{\'a}ke, including the Haudenosaunee and Anishinabeg peoples and the Kanien'keh{\'a}:ka Nation. The authors thank the anonymous referee for their constructive comments and insight. HB is grateful for support from the Natural Sciences and Engineering Research Council of Canada (NSERC) Alexander Graham Bell Canada Graduate Scholarship. HB and DH acknowledge funding from the NSERC Discovery Grant and the Canada Research Chairs (CRC) program. The scientific results reported in this article are based on observations made by the Chandra X-ray Observatory, the Spitzer Space Telescope, the Nuclear Spectroscopic Telescope Array (NuSTAR), the GRAVITY instrument on the Very Large Telescope,  and the Atacama Large Millimeter/submillimeter Array (ALMA). We thank the Chandra, Spitzer, NuSTAR, ALMA, and GRAVITY scheduling, data processing, and archive teams for making these observations possible. \\
\textit{Software}: \texttt{CIAO} \cite{2006SPIE.6270E..1VF}, \texttt{NumPy} \citep{SciPy}, \texttt{AstroPy} \citep{2018arXiv180102634T}, \texttt{matplotlib} \citep{Hunter:2007}, \texttt{bayesian blocks} \citep{2017ascl.soft04001W}, \texttt{zdcf} \citep{2013arXiv1302.1508A}, \texttt{flaremodel} \citep{Dallilar+2022} \\
\textit{Facilities}: Spitzer/IRAC, Chandra/ACIS, NuSTAR, VLTI/GRAVITY, ALMA

\bibliographystyle{aasjournal}
\bibliography{boyce_SgrA_IR-Xray_paper2}{}

\appendix

\section{SED Values}

\begin{table*}[h]
\caption{Values for the Sgr A* SED observed by coordinated ground-based and space-based observatories on 2019 July 18.}             
\label{tab:sed}      
\begin{tabularx}{\linewidth}{@{\extracolsep{\fill}}l c c c c c}
\hline
\multicolumn{1}{c}{} & \multicolumn{1}{c}{}& \multicolumn{2}{c}{$t=0$ minutes} & \multicolumn{2}{c}{$t=35$ minutes} \\
\cmidrule(rl){3-4} \cmidrule(rl){5-6}
\hline
Observatory & Frequency & Flux Density & $\nu$L$_{\nu}$ & Flux Density & $\nu$L$_{\nu}$\\ 
&[GHz]&[Jy]&[$\times10^{34}$erg\,s$^{-1}$]&[Jy]&[$\times10^{34}$erg\,s$^{-1}$]\\
\hline  
   ALMA\textsuperscript{a} & 340 & -- & -- & 2.6 $\pm$ 0.5 & 7.2 $\pm$ 1.4\\
   Spitzer/M-band & 6.7$\times10^{4}$ & 20.2 $\pm$ 1.0 $\times10^{-3}$ & 10.9 $\pm$ 0.6 & -- & -- \\
   GRAVITY/K-band & 1.4$\times10^{5}$ & 13.5 $\pm$ 0.9 $\times10^{-3}$ & 15.0 $\pm$ 1.0 & 0.8 $\pm$ 0.2 $\times10^{-3}$ & 0.9 $\pm$ 0.3\\
   GRAVITY/H-band & 1.9$\times10^{5}$ & 12.7 $\pm$ 1.4 $\times10^{-3}$ & 19.4 $\pm$ 2.2 & 1.0 $\pm$ 0.5 $\times10^{-3}$ & 1.5 $\pm$ 0.8\\
   Chandra & 6.8$\times10^{8}$ & 18.1 $\pm$ 4.7 $\times10^{-7}$ & 10.0 $\pm$ 2.6 & $<$ 7 $\times10^{-9}$ & $<0.3$\\
   NuSTAR & 1.5$\times10^{9}$ & 2.6 $\pm$ 0.7 $\times10^{-7}$  & 3.1 $\pm$ 0.9 & -- & -- \\
\hline
\end{tabularx}
\newline
\textsuperscript{a}After subtracting the $\sim$ 2 Jy contribution from the thermal component (grey line in Figure \ref{fig:sed})
\tablecomments{Frequencies for X-ray observatories reflect the central frequency of the keV energy band within the observation bin.}
\end{table*}

\section{Comparison with previous NIr/X-ray studies}\label{appendix}

Several other works have reported simultaneous X-ray and IR observations of Sgr A*. Some report simultaneity between the X-ray and IR peaks, but do not report a time frame within which that claim can be considered valid \citep{2006ApJ...644..198Y,2009ApJ...706..348Y,2011A&A...528A.140T}. Those that constrain timing between X-ray and IR activity \citep{2004A&A...427....1E,2006A&A...450..535E,2007ApJ...667..900H,2008A&A...479..625E,2009ApJ...698..676D,2012A&A...537A..52E,2017MNRAS.468.2447P,2007ApJ...667..900H,2012AJ....144....1Y} are plotted in the Appendix in Figure \ref{fig:xray_ir_timeLag} along with the \textit{Sptizer}-Chandra results of this campaign.

\begin{figure*}
\centering
\includegraphics[width=0.85\textwidth]{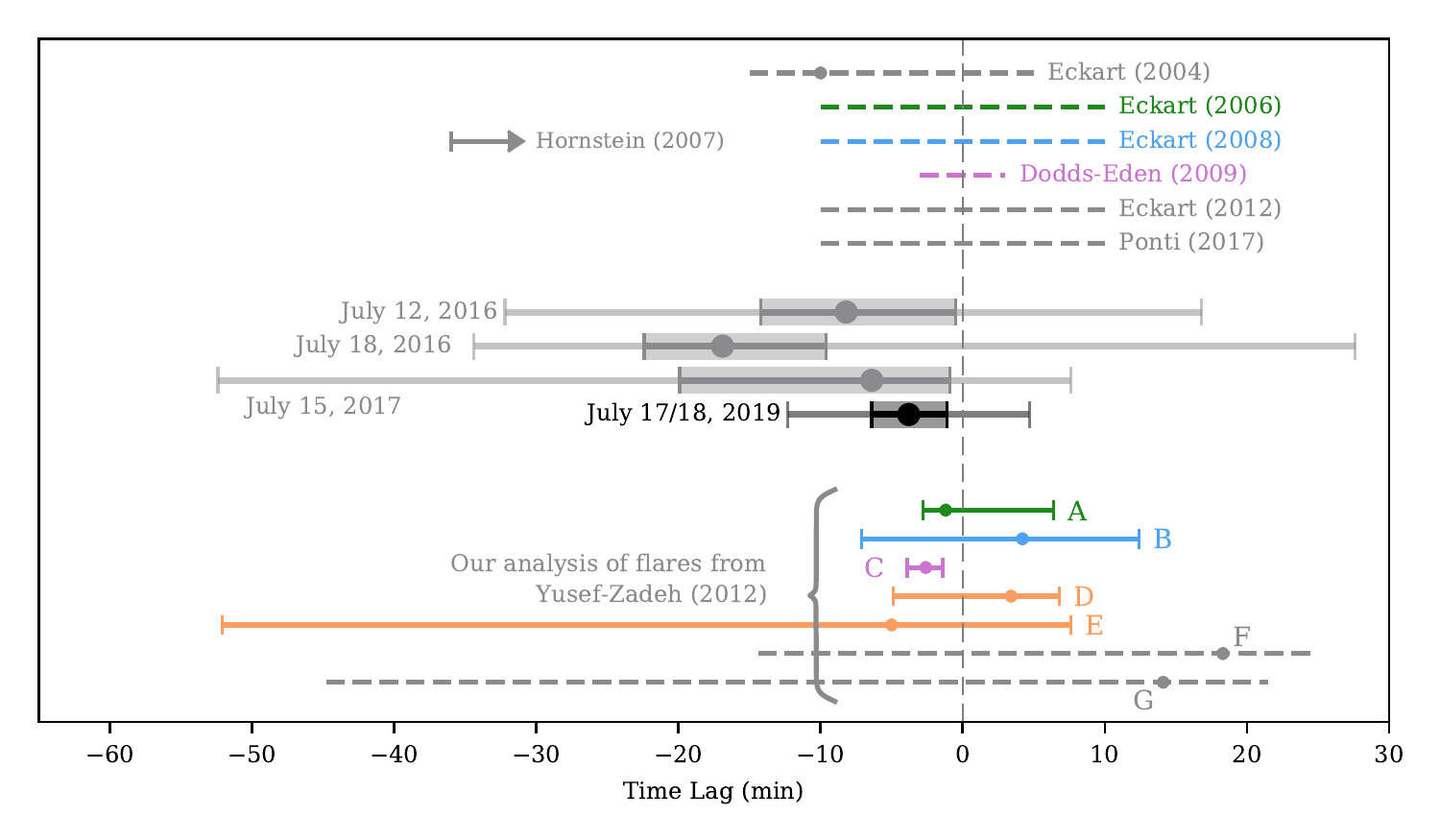}
\caption{Time lags between IR and X-ray flares as reported in this work and in the literature. Plotted with black circles are the time lags from the epochs in this Spitzer/Chandra campaign with significant X-ray and IR activity and their 68\% confidence intervals. Plotted in solid-grey are the updated results from re-analyzing data in \cite{2019ApJ...871..161B}. Regions marked with dashed lines come from works that describe the flares to be ``simultaneous to within $x$ minutes" but quote no uncertainties \citep{2004A&A...427....1E,2006A&A...450..535E,2007ApJ...667..900H,2008A&A...479..625E,2009ApJ...698..676D,2012A&A...537A..52E,2017MNRAS.468.2447P}. The upper limit from \cite{2007ApJ...667..900H} indicates an X-ray flare whose peak occurred 36 minutes before IR observations began. \cite{2012AJ....144....1Y} is the only work to report any correlation between the X-ray and IR with error bars. \cite{2019ApJ...871..161B} re-analyzed the seven flares presented in their work and we plot those results with 68\% confidence intervals here. Five of these flares come from previously reported data sets (color coded as green, blue, magenta and orange for \cite{2006A&A...450..535E}, \cite{2008A&A...479..625E}, \cite{2009ApJ...698..676D}, and \cite{2009ApJ...706..348Y} respectively) and two come from a previously un-reported data set (plotted in grey).}
\label{fig:xray_ir_timeLag}
\end{figure*}

\begin{table}[h]
\centering
\caption{Time Lags: Spitzer/Chandra Flares}
\label{tab:timeLags_old}
\begin{tabularx}{0.7\textwidth}{@{\extracolsep{\fill}}lcrrr}
	
	\toprule
	Date & time lag (min) & 68\% interval & 99.7\% interval \\
	\midrule
	2016 July 12 & $-13.5\substack{+5.2 \\ -5.1}$ & ($-18.6$,$-8.3$)  & ($-29.8$, $+2.8$) \\
	2016 July 18 & $-14.4\substack{+20.4 \\ -5.1}$ & ($-19.5$,$+6.0$) & ($-27.5$,$+18.6$) \\
	2017 July 15 & $-10.9\substack{+3.7 \\ -4.8}$ & ($-15.7$,$-7.2$) & ($-52.1$, $+0.4$) \\
	\textbf{*2019 July 18} & $-2.8\substack{+3.3 \\ -3.3}$ & ($-6.1$,$+0.5$) & ($-12.2$, $+6.7$) \\
	\bottomrule
	
\end{tabularx}

\hspace{-11cm}\textbf{*This work.} \\
\justify
\textit{Note}: Negative values mean X-ray leads IR.  Uncertainties on the time lag in the second column span the 68\% confidence interval on the \iterations MC runs. The second column displays the boundaries of this 68\% confidence interval, while the third column displays the 99.7\% confidence interval.

\end{table}

\end{document}